\documentclass[journal, romanappendices, final]{IEEEtran}
\usepackage{latexsym,amsfonts,amssymb,amsthm, setspace, verbatim,cite,mathrsfs,graphicx,bm,stfloats, hyperref, tikz,xcolor,soul}
\usepackage[cmex10]{amsmath}
\newcommand{\be}{\begin{equation}}
\newcommand{\ee}{\end{equation}}
\newcommand{\ben}{\begin{equation*}}
\newcommand{\een}{\end{equation*}}
\newcommand{\mc}{\mathcal}
\newtheorem{lem}{Lemma}

\newtheorem{thm}{Theorem}

\newtheorem{corr}{Corollary}

\newcommand{\e}{\epsilon}
\newcommand{\mcb}{\mathcal{B}_{M,L}}
\newcommand{\mbf}{\mathbf}
\newcommand{\bfs}{\mathbf{S}}
\newcommand{\bfsh}{\mathbf{\hat{S}}}

\newcommand{\abs}[1]{\lvert#1\rvert}
\newcommand{\norm}[1]{\lVert#1\rVert}
\newcommand{\expec}{\mathbb{E}}

\begin{document}
\title{Lossy Compression via Sparse Linear Regression: Computationally Efficient Encoding and Decoding}
\author{Ramji Venkataramanan,~\IEEEmembership{Member,~IEEE,}
Tuhin Sarkar,
and Sekhar Tatikonda,~\IEEEmembership{Senior Member,~IEEE}

\thanks{ This work was partially supported by NSF Grants CCF-1017744 and CCF-1217023. This paper was presented at the 2013 IEEE International Symposium on Information Theory.}%
\thanks{R.~Venkataramanan is with the Department of Engineering, University of Cambridge, Cambridge CB2 1PZ, UK (e-mail: ramji.v@eng.cam.ac.uk).}%
\thanks{T.~Sarkar is with the Department of Electrical Engineering, Indian Institute of Technology, Bombay
(e-mail: tuhin91@gmail.com).}
\thanks{S. Tatikonda is with the Department of Electrical Engineering, Yale University, New Haven CT 06511, USA (e-mail: sekhar.tatikonda@yale.edu).}
\thanks{Communicated by M.~Elad, Associate Editor for Signal Processing.}
}
\maketitle

\begin{abstract}
We propose computationally efficient encoders and decoders for lossy compression using a Sparse Regression Code. The codebook is defined by a design matrix and codewords are structured linear combinations of columns of this matrix. The proposed encoding algorithm  sequentially chooses columns of the design matrix to  successively approximate the source sequence. It is shown to achieve the optimal distortion-rate function for i.i.d Gaussian sources under the squared-error distortion criterion.  For a given rate, the parameters of the design matrix can be varied to trade off distortion performance with encoding complexity. An example of such a trade-off as a function of the block length $n$ is the following. With  computational resource (space or time) per source sample of $O((n/\log n)^2)$,  for a fixed distortion-level above the Gaussian distortion-rate function, the probability of excess distortion  decays exponentially in $n$. The Sparse Regression Code is  robust in the following sense: for any ergodic source, the proposed encoder achieves the optimal distortion-rate function of an i.i.d Gaussian source with the same variance. Simulations show that the encoder has good empirical performance, especially at low and moderate rates.
\end{abstract}


\section{Introduction}
\label{sec:intro}
\IEEEPARstart{D}{eveloping}   efficient codes for lossy compression at rates approaching the Shannon rate-distortion limit has long been one of the important goals of information theory. Efficiency is measured in terms of the storage complexity of the codebook as well the computational complexity of encoding and decoding. The Shannon-style i.i.d random codebook \cite{CoverThomas} has optimal performance in terms of the trade-off between distortion and rate as well as the error exponent\footnote{The error exponent of a compression code measures how fast the probability of excess distortion decays to zero with growing block length.} \cite{MartonRD74, IharaKubo00}. However, both the storage and computational complexity of this codebook  grow exponentially with the block length.

In this paper, we study a class of codes called Sparse Superposition or Sparse Regression Codes (SPARC) for lossy compression with the squared-error distortion criterion. We present computationally efficient encoding and decoding algorithms that provably attain the optimal rate-distortion function for i.i.d Gaussian sources.

The Sparse Regression codebook is constructed based on the statistical framework of high-dimensional linear regression, and was proposed recently by Barron and Joseph for communication over the AWGN channel at rates approaching the channel capacity \cite{AntonyML,AntonyFast}. The codewords are sparse linear combinations of columns of an $n \times N$ design matrix or `dictionary', where $n$ is the block-length and $N$ is a low-order polynomial in $n$. This structure enables the design of computationally efficient encoders based on sparse approximation ideas (e.g., \cite{MP93,BarronCDD08}). We propose one such encoding algorithm and analyze it performance.

SPARCs for lossy compression were first considered in \cite{KontSPARC} where some preliminary results were presented. The rate-distortion and error exponent performance of these codes under minimum-distance (optimal) encoding are characterized in a companion paper \cite{RVGaussianRD12}. The main contributions of this paper are the following.
\begin{itemize}
\item We propose a computationally efficient encoding algorithm for SPARCs which achieves the optimal distortion-rate function for i.i.d Gaussian sources with growing block length $n$.  The algorithm is based on successive approximation of the source sequence by columns of the design matrix. The parameters of the design matrix can be chosen to trade off performance with complexity. For example, one choice of parameters discussed in Section \ref{sec:main_result} yields an $n \times O(n^2)$ design matrix and per-sample encoding complexity proportional to $(\frac{n}{\log n})^2$. For this choice,  the probability of excess distortion for an i.i.d  Gaussian source (for a fixed distortion-level above the distortion-rate function) decays exponentially in $n$. To the best of our knowledge, this is the fastest proven rate of decay among lossy compressors with computationally feasible encoding and decoding.

\item With this encoding algorithm, SPARCs share the following robustness property of random i.i.d Gaussian codebooks \cite{Lapidoth97,SakMismatch1,SakMismatch2}: for a given rate $R$  nats, any ergodic source with variance $\sigma^2$ can be compressed with distortion close to the optimal i.i.d Gaussian distortion-rate function $\sigma^2 e^{-2R}$.

\item  The proposed encoding algorithm may be interpreted in terms of successive refinement \cite{EqCover91, Rimoldi94}. Letting $L=\frac{\log n}{n}$, one may interpret the algorithm as successively refining the source over $L$ stages, with rate $R/L$ in each stage. In other words, by successively refining the source over an asymptotically large number ($L$) of stages with  asymptotically small rate ($R/L$) in each stage, we attain the optimal Gaussian distortion-rate function with polynomial encoding complexity ($L^2$) and probability of excess distortion falling exponentially in $n$.

This successive refinement interpretation (discussed in Remark $7$ in Section \ref{sec:main_result}) is  of interest beyond the context of SPARCs, and could be used to develop computationally efficient lossy compression algorithms for general sources and distortion measures.
\end{itemize}

We remark that  for the proposed encoder with complexity that scales as a low-order polynomial in $n$, the gap between the typical realized distortion and the i.i.d Gaussian distortion-rate function is of the order of $\tfrac{\log \log n}{\log n}$. Designing feasible encoders with faster convergence to the rate-distortion function is an interesting open question, given the excellent error-exponent performance of SPARCs with optimal (minimum-distance) encoding \cite{RVGaussianRD12}.

The results of this paper  together with those in \cite{AntonyFast} show that Sparse Regression codes with computationally efficient encoders and decoders can be used for both source and channel coding at rates approaching the Shannon-theoretic limits. Further, the source and channel coding SPARCs can be nested to implement binning and superposition \cite{RVAller12}, which are essential ingredients of coding schemes for a large number of multi-terminal source and channel coding problems. Thus SPARCs can be used to build computationally efficient, rate-optimal codes for a variety of problems in network information theory.

We briefly review related work in developing computationally efficient codes for lossy compression. Gupta, Verd{\'u} and Weissman \cite{GuptaVerduWeiss} showed that the optimal rate-distortion function of memoryless sources can be approached by concatenating optimal codes over sub-blocks of length much smaller than the overall block length. Nearest neighbor encoding is used over each of these sub-blocks, which is computationally feasible due to their short length. For this scheme, it is not known how rapidly the probability of excess distortion decays to zero with the overall block length; the decay may be slow if the sub-blocks are chosen to be very short in order to keep the encoding complexity low. For sources with finite alphabet, various coding techniques have been proposed recently to approach the rate-distortion bound with computationally feasible encoding and decoding \cite{KontGioran,JalaliWeiss, GuptaVerdu09, WainManeva10,polarrd}. The rates of decay of the probability of excess distortion for these schemes vary, but in general they are slower than exponential in the block length.

The survey paper by Gray and Neuhoff \cite{GrayNeuhoff98} contains an extensive discussion of various compression techniques and their performance versus complexity trade-offs. These include scalar quantization with entropy coding, tree-structured vector quantization, multi-stage vector quantization, and trellis-coded quantization. Though these techniques have good empirical performance, they have not been shown to attain the optimal rate-distortion trade-off with computationally feasible encoders and decoders.  For an overview and comparison of these compression techniques, the reader is referred to \cite[Section V]{GrayNeuhoff98}. We remark that many of these schemes also use successive approximation ideas to reduce encoding complexity.
Lattice-based codes for lossy compression  \cite{ConSloane82,EyForney93,HamkZ02}  have a compact representation, i.e., low storage complexity. There are computationally  efficient quantizers for certain classes of lattice codes, but  the high-dimensional lattices needed to provably approach the rate-distortion bound have exponential encoding complexity in general \cite{Zamir02}.

The paper is organized as follows. Section \ref{sec:sparc} describes the construction of the sparse regression codebook. In Section \ref{sec:alg_desc}, we describe the encoding algorithm, followed by a heuristic explanation of why it attains the Gaussian distortion-rate limit. Section \ref{sec:main_result} contains the main result of the paper, a characterization of the compression performance of SPARCs with the proposed encoding algorithm. Various remarks are also made regarding the performance-complexity tradeoff,  gap from the optimal distortion-rate limit, the successive refinement interpretation etc. Section \ref{sec:main_result} also contains simulation results illustrating the distortion-rate performance. The proof of the main result is given in Section \ref{sec:proof}, and Section \ref{sec:conc} concludes the paper.

\emph{Notation}: Upper-case letters are used to denote random variables, lower-case for their realizations,  and bold-face letters for random vectors and matrices. $\mc{N}(\mu,\sigma^2)$ denotes the Gaussian distribution with mean $\mu$ and variance $\sigma^2$.
All vectors  have length $n$. The  source sequence  is denoted by $\bfs \triangleq (S_1, \ldots, S_n)$, and the reconstruction sequence by $\bfsh \triangleq (\hat{S}_1, \ldots, \hat{S}_n)$.
$\norm{\mathbf{X}}$ denotes the $\ell_2$-norm of vector $\mathbf{X}$, and
$\abs{\mathbf{X}} =  \norm{\mathbf{X}} / \sqrt{n}$ is the normalized version. $\langle \mathbf{a}, \mathbf{b} \rangle = \sum_i a_i b_i$ denotes the Euclidean inner  product between vectors $\mbf{a}$ and $\mbf{b}$.  $f(x)=o(g(x))$ means $\lim_{x \to \infty} f(x)/g(x) =0$; $f(x)=\Theta(g(x))$ means $f(x)/g(x)$ asymptotically lies in an interval $[\kappa_1,\kappa_2]$ for some constants $\kappa_1,\kappa_2>0$.  All logarithms are with base $e$ unless otherwise mentioned, and rate is measured in nats.

\section{The Sparse Regression Codebook} \label{sec:sparc}
\begin{figure}[t]
\centering
\includegraphics[width=3.5in]{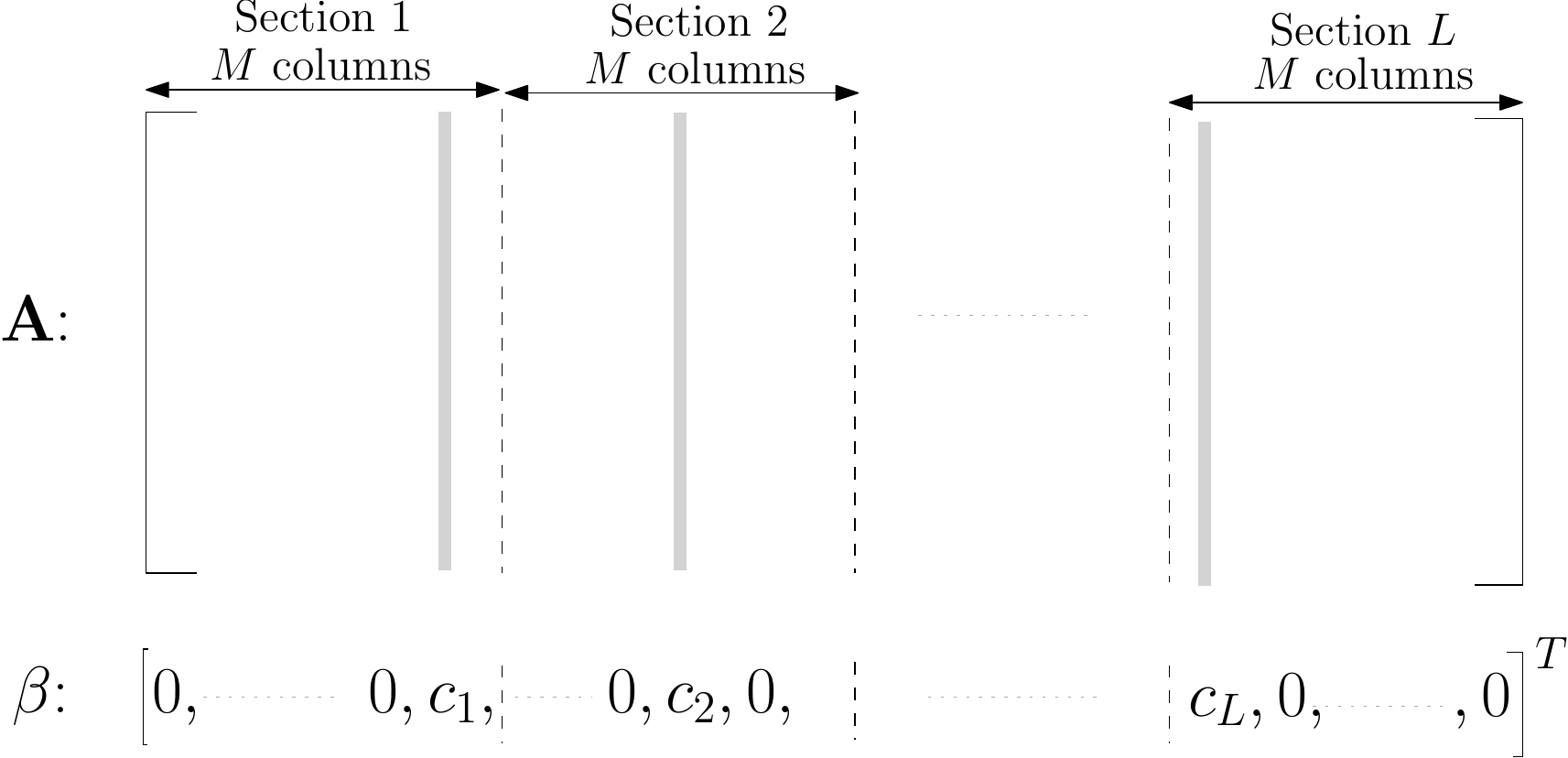}
\caption{\small{$\mathbf{A}$ is an $n \times ML$ matrix and $\beta$ is a $ML \times 1$ vector. The positions of the non-zeros in $\beta$  correspond to the gray columns of $\mathbf{A}$ which combine to form the codeword $\mathbf{A}\beta$.}}
\vspace{-5pt}
\label{fig:sparserd}
\end{figure}

A sparse regression code (SPARC) is defined in terms of a design matrix $\mathbf{A}$ of dimension $n \times ML$ whose entries are i.i.d. $\mathcal{N}(0,1)$, i.e., independent zero-mean Gaussian random variables with unit variance. Here $n$ is the block length and $M$ and $L$ are integers whose values will be specified shortly in terms of $n$ and the rate $R$.  As shown in Fig. \ref{fig:sparserd}, one can think of the matrix $\mathbf{A}$ as composed of $L$ sections with $M$ columns each. Each codeword is a linear combination of $L$ columns, with one column from each section.
Formally, a codeword can be expressed as  $\mathbf{A} \beta$, where $\beta$ is  an $ML \times 1$ vector $(\beta_1, \ldots, \beta_{ML})$ with the following property:  there is exactly one non-zero $\beta_j$ for  $1 \leq j \leq M$, one non-zero $\beta_j$ for $M+1 \leq j \leq 2M$, and so forth.  The non-zero value of $\beta$ in section $i$  is set to $c_i$ where the value of  $c_i$ will be specified in the next section.  Denote the set of all $\beta$'s that satisfy this property by $\mcb$.

Since there are $M$ columns in each of the $L$ sections, the total number of codewords is $M^L$. To obtain a compression rate of $R$ nats/sample, we therefore need
\be
M^L = e^{nR} \quad \text{ or } \quad L \log M = nR
\label{eq:ml_nR}
\ee

\emph{Encoder}: This is defined by a mapping $g: \mathbb{R}^n \to \mcb$. Given the source sequence $\bfs$ and target distortion $D$, the encoder attempts to find a $\hat{\beta} \in \mcb$ such that
 \[  \norm{\bfs - \mathbf{A}\hat{\beta}}^2 \leq D.\]
 If such a codeword is not found, an error is declared. In the next section, we present a computationally efficient encoding algorithm and characterize its performance in  Section \ref{sec:main_result}.

\emph{Decoder}: This is a mapping $h: \mcb \to \mathbb{R}^n$. On receiving $\hat{\beta} \in \mcb$ from the encoder, the decoder produces reconstruction\footnote{In the remainder of the paper, we will often refer to $\hat{\beta}$ as the codeword though, strictly speaking, $\mathbf{A} \hat{\beta}$ is the actual codeword.} $h(\hat{\beta}) = \mathbf{A}\hat{\beta}$.

\emph{Storage Complexity}: The storage complexity of the dictionary is proportional to $nML$. There are several choices for the pair $(M,L)$ which satisfy \eqref{eq:ml_nR}. For example, $L=1$ and $M=e^{nR}$ recovers the Shannon-style random codebook in which the number of columns in the dictionary $\mathbf{A}$ is $e^{nR}$, i.e., the storage complexity is exponential in $n$. For our constructions in Section \ref{subsec:sim}, we will choose $M$ to be a low-order polynomial in $n$. This implies that $L$ is $\Theta\left(\frac{n}{\log n}\right)$, and the number of columns $ML$ in the dictionary  is a low-order polynomial in $n$.
This reduction in storage complexity can be harnessed to develop computationally efficient encoders for the sparse regression code. We emphasize that the results presented here hold for any choice of $(M,L)$ satisfying \eqref{eq:ml_nR}; the choice of $(M,L)$ above offers a good trade-off between complexity and error performance.

\section{Computationally Efficient Encoding Algorithm} \label{sec:alg_desc}

The source sequence $\bfs$ is generated by an ergodic source with mean $0$ and variance $\sigma^2$.

The SPARC is defined by the $n \times ML$ design matrix $\mathbf{A}$. The $j$th column of $\mathbf{A}$ is denoted $\mbf{A}_j$, $1 \leq j \leq ML$.
$\{c_i\}_{i=1}^L$, the non-zero values of $\beta$, are chosen to be
\be
c_i = \sqrt{\frac{2R \sigma^2}{L} \left( 1- \frac{2R}{L} \right)^{i-1}}, \quad i=1, \ldots, L.
\label{eq:ci_def}
\ee
Given source sequence $\bfs$, the encoder determines $\hat{\beta} \in \mcb$ according to the following algorithm.

\emph{Step} $0$: Set $\mbf{R}_0 = \bfs$.

\emph{Step} $i$, $i=1, \ldots, L$: Pick
\be  m_i =  \underset{j: \ (i-1)M < \; j \; \leq iM }{\operatorname{argmax}}
\left\langle \mbf{A}_j,  \frac{\mbf{R}_{i-1}}{\norm{\mbf{R}_{i-1}}} \right\rangle. \label{eq:max_stepi} \ee
Set
\begin{equation}
\mathbf{R}_{i} = \mathbf{R}_{i-1} - c_i \mathbf{A}_{m_i},
\label{eq:gen_residue}
\end{equation}
where $c_i$ is given by \eqref{eq:ci_def}.

\emph{Step} $L+1$:  The codeword $\hat{\beta}$  has  non-zero values in positions $m_i, \ 1 \leq i \leq L$. The value of the non-zero in section $i$ given by $c_i$.

The algorithm chooses  the $m_i$'s in a greedy manner - section by section - to minimize a `residue' in each step. In Section \ref{subsec:heur}, we give a non-rigorous explanation of why the  algorithm succeeds (with high probability)  in finding a codeword within distortion $D$ of a typical source sequence, for rates $R$ larger than $R^*(D)$, the i.i.d. Gaussian rate-distortion function. The formal performance analysis is contained  in Sections \ref{sec:main_result} and \ref{sec:proof}.

\subsection{Computational Complexity}
The encoding algorithm consists of $L$ stages, where each stage involves computing $M$ inner products followed by finding the maximum among them. Therefore the number of operations per source sample is proportional to $ML$. If we choose $M=L^b$ for some $b>0$,  \eqref{eq:ml_nR} implies that  $L=\Theta\left( \frac{n}{\log n}\right)$, and the number of operations per source sample is of the order  $\left( n/ \log n \right)^{b+1}$. We also note that due to the sequential nature of the algorithm, it is enough to do a single pass on the codebook. At any time, in addition to the current residue, only one length $n$ column of the matrix needs to be kept in memory.

When we have several source sequences to be encoded in succession, the encoder can have the following pipelined architecture. There are $L$  modules: the first module computes the inner product of the source sequence  with each column in the first section of $\mathbf{A}$ and determines the maximum; the second module computes the inner product of the first-step residual vector with each column in the second section of $\mathbf{A}$, and so on. Each module has $M$ parallel units; each unit consists of a multiplier and an accumulator to compute an inner product in a pipelined fashion. After an initial delay of $L$ source sequences, all the modules work simultaneously. This encoder architecture requires computational space (memory) of the order $nLM$ and has constant computation time per source symbol.

The code structure automatically yields low decoding complexity. The encoder can represent the chosen $\beta$ with $L$ binary sequences of $\log_2 M$ bits each. The $i$th binary sequence indicates the position of the non-zero element in section $i$. Hence the decoder complexity corresponding to locating the $L$ non-zero elements using the received bits is $L \log_2 M$, which is $O(1)$ per source sample. Reconstructing the codeword then requires $L$ additions per source sample.

\subsection{Heuristic derivation of the algorithm} \label{subsec:heur}

In this section, we  present a non-rigorous analysis of the proposed encoding algorithm based on the following observations.
\begin{enumerate}

\item For $1 \leq j \leq ML$, $\abs{\mbf{A}_j}^2$ is approximately equal to $1$  when $n$ is large. This is due to the law of large numbers since each $\abs{\mbf{A}_j}^2$ is the normalized sum of squares of $n$ i.i.d. $\mc{N}(0,1)$ random variables.

\item Similarly, $\abs{\mathbf{S}}^2$ is approximately equal to $\sigma^2$ for large $n$ due to the law of large numbers.

\item If $X_1,X_2\ldots, X_M$ are i.i.d. $\mc{N}(0,1)$ random variables, then $\max\{ X_1,\ldots, X_M\}$ is approximately equal to $\sqrt{2\log M}$ for large $M$ \cite{DavidNagaraja}.
\end{enumerate}
The deviations of these quantities from their typical values above are precisely characterized in Section \ref{sec:proof}. 

We begin with the following lemma about projections of i.i.d. Gaussian random vectors. 
\begin{lem}
Let $\mbf{A}_1, \ldots, \mbf{A}_N$ be $N$ mutually independent random vectors with i.i.d. $\mc{N}(0,1)$ components. 
Then, for any random vector $\mbf{R}$ which is independent of the collection $\{\mbf{A}_j\}_{j=1}^N$ and has support on the unit sphere in $\mathbb{R}^n$ , the inner products
\[ T_j \triangleq \left\langle  \mbf{A}_j,  \mbf{R} \right\rangle, \quad j=1,\ldots,N \]
are i.i.d. $\mc{N}(0,1)$ random variables that are independent of $\mbf{R}$.
\label{lem:inner_prod}
\end{lem}
\begin{IEEEproof}
In Appendix \ref{app:inner_prod}. 
\end{IEEEproof}

We note that the lemma allows $\mbf{R}$ to be a deterministic vector. 

\emph{Step} $1$: Consider the statistic
\be T^{(1)}_{j} \triangleq \left\langle  \mbf{A}_j, \frac{\mbf{R}_{0}}{\norm{\mbf{R}_{0}}} \right\rangle,  \quad  1 \leq  j  \leq M. \ee
Note that $\mbf{R}_0 = \mbf{S}$ is independent of each $\mbf{A}_j$, which are random vectors with $\mc{N}(0,1)$ components. 
Therefore, by Lemma \eqref{lem:inner_prod},  $T^{(1)}_{j}, \, 1 \leq j \leq M$ are i.i.d. $N(0,1)$ random variables. Hence
\be
\max_{1 \leq j \leq M} T^{(1)}_{j} = \left\langle  \mbf{A}_{m_1}, \frac{\mbf{R}_{0}}{\norm{\mbf{R}_{0}}}  \right\rangle \approx \sqrt{2 \log M}.
\label{eq:max_T1j}
\ee
From \eqref{eq:gen_residue}, the normalized norm of the residue $\mathbf{R}_1$ can be expressed as
\be
\begin{split}
\abs{\mathbf{R}_1}^2 & = \abs{\mathbf{R}_0}^2 + c_1^2 \abs{\mathbf{A}_{m_1}}^2  - \frac{2c_1}{n} \langle \mathbf{A}_{m_1}, \mathbf{R}_0 \rangle \\
& = \abs{\mathbf{R}_0}^2 + c_1^2 \abs{\mathbf{A}_{m_1}}^2  - \frac{2 c_1 \norm{\mathbf{R}_0}}{n}
\left\langle \mathbf{A}_{m_1}, \frac{\mathbf{R}_0}{\norm{\mathbf{R}_0}} \right\rangle\\
& \stackrel{(a)}{\approx} \abs{\mathbf{R}_0}^2  +  c_1^2 - \frac{2 c_1 \norm{\mathbf{R}_0}}{n}\sqrt{2\log M} \\
& \stackrel{(b)}{\approx} \sigma^2 + c_1^2  - \frac{2 c_1 \sigma}{\sqrt{n}}\sqrt{2\log M} \\
& \stackrel{(c)}{=} \sigma^2\left( 1 - \frac{2R}{L}\right).
\end{split}
\label{eq:R0_R1}
\ee
In the chain above $(a)$ and $(b)$ follow from \eqref{eq:max_T1j} and the three observations listed at the beginning of this subsection. $(c)$ follows by substituting for $c_1$ from \eqref{eq:ci_def} and for $n$ from \eqref{eq:ml_nR}.

\emph{Step $i$, $i=2,\ldots,L$}: We show  that if
\[ \abs{\mathbf{R}_{i-1}}^2 \approx \sigma^2\left( 1 - \frac{2R}{L}\right)^{i-1}, \]
then
\be \abs{\mathbf{R}_{i}}^2 \approx \sigma^2\left( 1 - \frac{2R}{L}\right)^{i}.  \label{eq:Res_i}\ee
We already showed that \eqref{eq:Res_i} is true for $i=1$.

For each $j \in \{ (i-1)M+1, \ldots, iM \}$, consider the statistic
\be T^{(i)}_{j} \triangleq \left\langle  \mbf{A}_j,  \frac{\mbf{R}_{i-1}}{\norm{\mbf{R}_{i-1}}} \right\rangle.   \label{eq:stat_Tij}\ee
Note that $\mbf{R}_{i-1}$ is {independent} of $\mbf{A}_j$ because $\mbf{R}_{i-1}$ is a \emph{function} of the source sequence $\mbf{S}$ and the columns $\{\mbf{A}_j\}, \ 1\leq j \leq (i-1)M$, which are all independent of  $\mbf{A}_j$ for $j \in \{ (i-1)M+1, \ldots, iM \}$.  Therefore, by Lemma \eqref{lem:inner_prod}, the $T^{(i)}_{j}$'s are i.i.d. $\mc{N}(0,1)$ random variables for $j \in \{ (i-1)M+1, \ldots, iM\}$. Hence, we have
\be
\max_{(i-1)M+1 \, \leq j \, \leq iM} \; T^{(i)}_{j} = \left\langle  \mbf{A}_{m_i}, \frac{\mbf{R}_{i-1}}{\norm{\mbf{R}_{i-1}}} \right\rangle \approx \sqrt{2 \log M}.
\label{eq:max_Tij}
\ee
From \eqref{eq:gen_residue}, we have
\be
\begin{split}
&\abs{\mathbf{R}_i}^2   = \abs{\mathbf{R}_{i-1}}^2 + c_i^2 \abs{\mathbf{A}_{m_i}}^2  - \frac{2 c_i \norm{\mathbf{R}_{i-1}}}{n}
\left\langle \mathbf{A}_{m_i}, \frac{\mathbf{R}_{i-1}}{\norm{\mathbf{R}_{i-1}}} \right\rangle\\
& \stackrel{(a)}{\approx} \abs{\mathbf{R}_{i-1}}^2  +  c_i^2 - \frac{2 c_i \norm{\mathbf{R}_{i-1}}}{n}\sqrt{2\log M} \\
& \stackrel{(b)}{\approx} \sigma^2\left( 1 - \frac{2R}{L}\right)^{i-1} + c_i^2 - \frac{2 c_i \sigma \sqrt{\left( 1 - \frac{2R}{L}\right)^{i-1}}}{\sqrt{n}}\sqrt{2\log M} \\
& \stackrel{(c)}{=} \sigma^2\left( 1 - \frac{2R}{L}\right)^i.
\end{split}
\label{eq:Ri1_Ri}
\ee
As before, $(a)$ and $(b)$ follow from \eqref{eq:max_Tij} and the three observations listed at the beginning of this subsection. $(c)$ holds by substituting for $c_i$ from \eqref{eq:ci_def} and for $n$ from \eqref{eq:ml_nR}. It can be verified that the chosen value of $c_i$ minimizes the third line in \eqref{eq:Ri1_Ri}.

Therefore, the residue when the algorithm terminates after Step $L$ is
\be
\begin{split}
\abs{\mathbf{R}_L}^2 = \abs{\bfs - \mathbf{A}\hat{\beta}}^2 & \approx  \sigma^2\left( 1 - \frac{2R}{L}\right)^L  {\leq} \ \sigma^2 e^{-2R}
\end{split}
\ee
where we have used the inequality $(1+x) \leq e^{x}$ for  $x \in \mathbb{R}$.

Thus the encoding algorithm picks a codeword $\hat{\beta}$ that yields squared-error distortion approximately equal to $\sigma^2 e^{-2R}$, the Gaussian distortion-rate function at rate $R$. Making the arguments above rigorous involves bounding the deviation of the residual distortion each stage from its typical value.

\section{Main Result} \label{sec:main_result}

\begin{thm}
Consider a length $n$  source sequence $\bfs$ generated by an ergodic source with mean $0$ and variance $\sigma^2$.
Let $\delta_0, \delta_1, \delta_2$ be any positive constants such that
\be \Delta \triangleq \delta_0 + 5R (\delta_1 + \delta_2) < \frac{1}{2}. \label{eq:del0del1del2}\ee
Let $\mathbf{A}$ be an $n \times ML$ design matrix  with i.i.d. $\mc{N}(0,1)$ entries and $M,L$ satisfying \eqref{eq:ml_nR}. On the SPARC defined by $\mathbf{A}$, the proposed encoding algorithm produces a codeword $\mbf{A}\hat{\beta}$ that satisfies the following for  sufficiently large $M,L$.
\be P\left( \; \abs{\bfs - \mbf{A}\hat{\beta}}^2 \;  >  \; \sigma^2 e^{-2R}(1 + e^R \Delta)^2  \; \right) < p_0+ p_1 + p_2  \label{eq:p_err_ub} \ee
where
\begin{equation}
\begin{split}
p_0 & = P\left( \;  \left| \frac{\abs{\bfs}}{\sigma} -1 \right| > \delta_0 \; \right), \\
p_1 & = 2ML\cdot \exp\left( -n{\delta_1^2}/{8}\right),\\
p_2 & = \left( \frac{M^{2\delta_2}}{8 \log M} \right)^{-L}.
\end{split}
\label{eq:p0p1p2}
\end{equation}
\label{thm:rd_feasible}
\end{thm}
\begin{corr}
If the source sequence $\mathbf{S}$ is generated according to an i.i.d Gaussian distribution $\mc{N}(0, \sigma^2)$,  then the SPARC with $M = L^b$,
$b >0$ attains the optimal distortion-rate function $D^*(R) = \sigma^2 e^{-2R}$ with the proposed encoder. Further, for any fixed distortion-level above $D^*(R)$, the probability of excess distortion decays exponentially with the block length $n$ for sufficiently large $n$.
\label{corr:Gauss}
\end{corr}
\begin{IEEEproof}
For a fixed distortion-level $\sigma^2 e^{-2R} + \gamma$ with $\gamma >0$, we can find  $\Delta >0$ such that
\be
\sigma^2 e^{-2R} + \gamma = \sigma^2 e^{-2R}(1 + e^R \Delta)^2,
\ee
or
\be
 \gamma = \sigma^2 \Delta^2 + 2 \Delta e^R \sigma^2.
\label{eq:gam_Delta}
\ee
Without loss of generality, we may assume that $\gamma$ is small enough that  $\Delta$ satisfying \eqref{eq:gam_Delta} lies in the interval $(0, \tfrac{1}{2})$.   For any positive constants $\delta_0, \delta_1, \delta_2$ chosen to satisfy \eqref{eq:del0del1del2}, Theorem \ref{thm:rd_feasible} implies that
\be
P\left( \abs{\bfs - \mbf{A}\hat{\beta}}^2 \;  >  \; \sigma^2 e^{-2R} + \gamma \right) < p_0 + p_1 + p_2,
\label{eq:excess_gam}
\ee
where $p_0,p_1, p_2$ are given by \eqref{eq:p0p1p2}. We now obtain upper bounds for $p_0, p_1, p_2$.

 For an i.i.d Gaussian source, $\norm{\bfs}^2$ is the sum of the squares of $n$ i.i.d $\mc{N}(0, \sigma^2)$ random variables.
Using a Chernoff bound on the probability of the events $\{ \norm{\bfs}^2 > n\sigma^2(1+ \delta_0)\}$ and  $\{ \norm{\bfs}^2 < n\sigma^2(1 - \delta_0)\}$, we obtain
 \be p_0 <  2 \exp(- 3n \delta_0^2/4) \label{eq:p0_bound} \ee

 When $M = L^b$, \eqref{eq:ml_nR} implies that $L=\Theta\left( n/ \log n \right)$. Therefore $ML = L^{b+1}$ grows polynomially in $n$,  and the term $p_1$
in  \eqref{eq:p0p1p2} can be  expressed as
\be
p_1 = \exp\left(-n\left(\tfrac{\delta_1^2}{8} - O( \tfrac{\log n}{n}) \right)\right)
\label{eq:p1_bound}
\ee
From \eqref{eq:p0p1p2}, $p_2$ can be expressed as
\be
\begin{split}
-  \frac{1}{n} \log p_2& = \frac{2 \delta_2 L \log M}{n} -  \frac{L \log (8 \log M)}{n} \\
& {=} 2 R \delta_2  -  \frac{ R \log (8 \log M)}{\log M} \\
& {=} 2 R \delta_2 -  O\left( \frac{ \log \log n}{\log n} \right).
\end{split}
\label{eq:p2_bounda}
\ee
In \eqref{eq:p2_bound}, the first equality is obtained from \eqref{eq:ml_nR}, while the second holds because $M = L^b = O((n/\log n)^{b})$. Hence
\be
 p_2 = \exp\left( -n \left( 2 \delta_2 R - O\left( \tfrac{ \log \log n}{\log n} \right)\right) \right).
 \label{eq:p2_bound}
\ee
Using \eqref{eq:p0_bound}, \eqref{eq:p1_bound} and \eqref{eq:p2_bound} in \eqref{eq:excess_gam}, we see that for any fixed distortion-level $D^*(R) + \gamma$, the probability of excess distortion decays exponentially in $n$ when $n$ is sufficiently large.

\end{IEEEproof}
\emph{Remarks}:
\begin{enumerate}
\item
The probability measure in \eqref{eq:p_err_ub} is over the space of source sequences and design matrices. The codeword $\hat{\beta}$ is a deterministic function of the source sequence $\bfs$ and design matrix $\mbf{A}$.

\item Ergodicity of the source is only needed to ensure that $p_0 \to 0$ as $n \to \infty$ (at a rate depending only on the source distribution).

 \item For a given rate $R$, Theorem \ref{thm:rd_feasible} guarantees that the  proposed  encoder achieves a squared-error distortion close to $D^*(R)= \sigma^2 e^{-2R}$ for  all ergodic sources with variance $\sigma^2$. This complements results along the same lines by Sakrison and Lapidoth \cite{Lapidoth97,SakMismatch1,SakMismatch2} for  Gaussian random codebooks (i.i.d codewords) with minimum-distance encoding.  Lapidoth \cite{Lapidoth97} also showed that for any ergodic source  of a given variance, one cannot attain a squared-error distortion smaller than the $D^*(R)$ using a Gaussian random codebook.
\item \emph{Gap from $D^*(R)$}: To achieve distortions close to the $D^*(R)$ with high probability, we need $p_0,p_1,p_2$ to all go to $0$.
In particular, for $p_2 \to 0$ with growing $L$, from \eqref{eq:p0p1p2}  we require that
$ {M^{2\delta_2}} > {8 \log M}$.
Or,
\be
\delta_2 > \frac{\log \log M}{2 \log M} + \frac{\log 8}{2 \log M}.
\label{eq:delta_min}
\ee
To approach $D^*(R)$, note that we need $n, L, M$ to all go to $\infty$ while satisfying \eqref{eq:ml_nR}: $n,L$
for the probability of error in  \eqref{eq:p0p1p2} to be small, and $M$ in order to allow $\delta_2$ to be small according to \eqref{eq:delta_min}.
When $L, M$ grow polynomially in $n$, \eqref{eq:delta_min} dictates how small $\Delta$ can be:  the distortion is  $\Theta\left( \frac{\log \log M}{\log M} \right)$ higher than the optimal value $D^*(R)=\sigma^2 e^{-2R}$.

\end{enumerate}

\subsection{Performance versus Complexity Trade-off}
The storage complexity of the SPARC is proportional to $nML$, the number of entries in the design matrix. Recall that the computational complexity of the encoding algorithm is $\Theta(ML)$ operations per source sample. The performance of the algorithm improves as $M, L$ increase, both in terms of the gap from the optimal distortion \eqref{eq:delta_min} and  the probability of error \eqref{eq:p0p1p2}.  Let us  consider a few illustrative cases.

\begin{itemize}
\item Choosing $M=L^b$ for some $b>0$ yields $L = \Theta \left(\frac{n}{\log n}\right)$. Hence the per-sample computational complexity is $\Theta\left(({n}/{\log n})^{b+1}\right)$ and the gap from $D^*(R)$ governed by \eqref{eq:delta_min} is of the order of
$ \frac{\log \log n}{b \log n}$.  For our simulations described in the next sub-section, we choose $b=3$ and $L \in [50,100]$.

\item If we choose $M = \kappa \log n$ for $\kappa >0$, \eqref{eq:ml_nR} implies that $L = \frac{nR}{\log (\kappa \log n)}$. The per-sample computational complexity is $\Theta \left( \frac{n \log n}{\log \log n}\right)$,  lower than the previous case. However, the gap $\delta_2$ from  \eqref{eq:delta_min} is approximately $\frac{ \log \log  \log n}{ \log \log n}$, i.e., the convergence to $D^*(R)$ with $n$ is much slower.

\item At the other extreme, consider the Shannon codebook with $L=1, M=e^{nR}$. In this case, the SPARC consists of only one section and the proposed algorithm is essentially minimum-distance encoding. The computational complexity is $O(e^{nR})$ (exponential), while the gap $\delta_2$ from \eqref{eq:delta_min} is approximately $\frac{\log n}{n}$. The gap $\Delta$ from $D^*(R)$ is now dominated by $\delta_0$ and $\delta_1$ which are $\Theta(1/\sqrt{n})$, consistent with the results in \cite{SakrisonFin, IngberKochman, KostinaV12}.\footnote{ For $L=1$,  the factor $ML$ that multiplies the exponential term in $p_2$ can be eliminated via a sharper analysis.}
\end{itemize}

To achieve a distortion gap $\delta$ from $D^*(R)$, \eqref{eq:delta_min} indicates that $M$ has to be of the order of $e^{1/\delta}$, i.e., the complexity is exponential in 
$\tfrac{1}{\delta}$. Designing feasible encoders whose complexity grows polynomially with $\tfrac{1}{\delta}$ is an important open question. In terms of the block length $n$, such an encoder would achieve a distortion within  $O(n^{-\gamma})$ of $D^*(R)$ for some $\gamma \in (0, 0.5)$, and would have complexity growing polynomially in $n$.

\subsection{Successive Refinement Interpretation}
The proposed encoding algorithm may be interpreted in terms of successive refinement source coding \cite{EqCover91, Rimoldi94}. We can think of each section of the design matrix $\mbf{A}$ as a lossy codebook of rate $R/L$. For each section $i$, $i=1, \ldots, L$, the residue $\mbf{R}_{i-1}$ acts as the `source' sequence, and the algorithm attempts to find the column \emph{within} the section that minimizes the distortion. The distortion after Section $i$  is the variance of the residue $\mbf{R}_i$; this residue  acts as the source sequence for Section $i-1$.  Recall  that the minimum mean-squared distortion achievable with a Gaussian codebook  at rate $R/L$ is  \cite{Lapidoth97}
\be
\begin{split}
D^*_i & = \abs{\mbf{R}_{i-1}}^2 \exp(-2R/L)  \\
& \approx  \abs{\mbf{R}_{i-1}}^2 \left( 1 - \frac{2R}{L} \right), \quad  \text{ for } R/L \ll 1. 
\end{split}
\label{eq:Di_star}
\ee
This minimum distortion can be attained with a codebook with elements chosen i.i.d $\mc{N}(0, \abs{\mbf{R}_{i-1}}^2 - D^*_i)$. From \eqref{eq:ci_def}, recall that the codeword variance in section $i$ of the codebook is
\be
 c_i^2 = \frac{2R \sigma^2}{L} \left( 1- \frac{2R}{L} \right)^{i-1} \approx \abs{\mbf{R}_{i-1}}^2 - D^*_i,
\ee
    where the approximate equality follows from \eqref{eq:Di_star} and \eqref{eq:Res_i}. Therefore, the typical value of the distortion in Section $i$ is close to $D^*_i$ since the algorithm is equivalent to minimum-distance encoding within each section. However, since the rate $R/L$ is infinitesimal, the deviations from $D^*_i$  in each section can be quite significant. 
     Despite this,  when the number of sections $L$ is large the \emph{final} distortion $\abs{\mbf{R}_L^2}$ is close to the typical value
    $\sigma^2 e^{-2R}$. The proof of Corollary $1$  implies that  the probability that the final distortion  is greater than $\sigma^2 e^{-2R} + \gamma$  falls \emph{exponentially} in $n$, for any fixed $\gamma$. This holds for any source whose second moment satisfies a large deviations property, i.e.,  $P(\abs{\mbf{S}}^2  > \sigma^2 + \delta)$ decays exponentially in $n$ for any
    fixed $\delta >0$.

  The redundancy of a code is the gap between its expected distortion and the Shannon distortion-rate function. An upper bound is derived  in \cite{yangZ06} on the redundancy of successive refinement codes. This  bound grows linearly in the number of stages $L$.  Though our results bound the probability of excess distortion rather than the redundancy, they suggest that the upper bound  in \cite{yangZ06} may not be tight. Determining the redundancy of the proposed SPARC encoder is an interesting open question.

    We emphasize that the successive refinement interpretation is only true for the proposed encoding algorithm, and is \emph{not} an inherent feature of the  sparse regression codebook. In particular, an important direction for future work is to design encoding algorithms with faster convergence to $D^*(R)$ while still having complexity that is polynomial in $n$.

\begin{figure}
\includegraphics[width=3.45in]{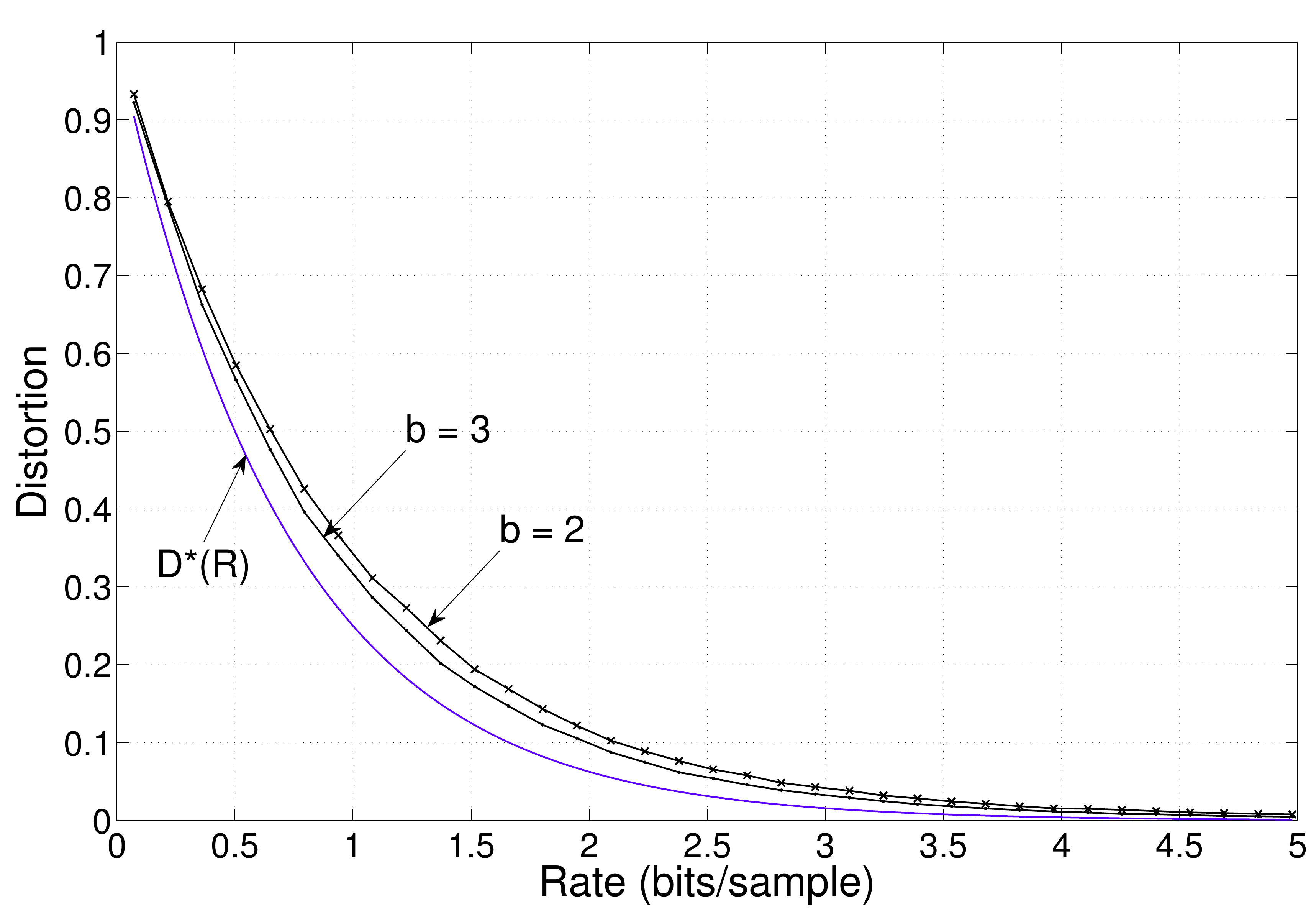}
\includegraphics[width=3.44in]{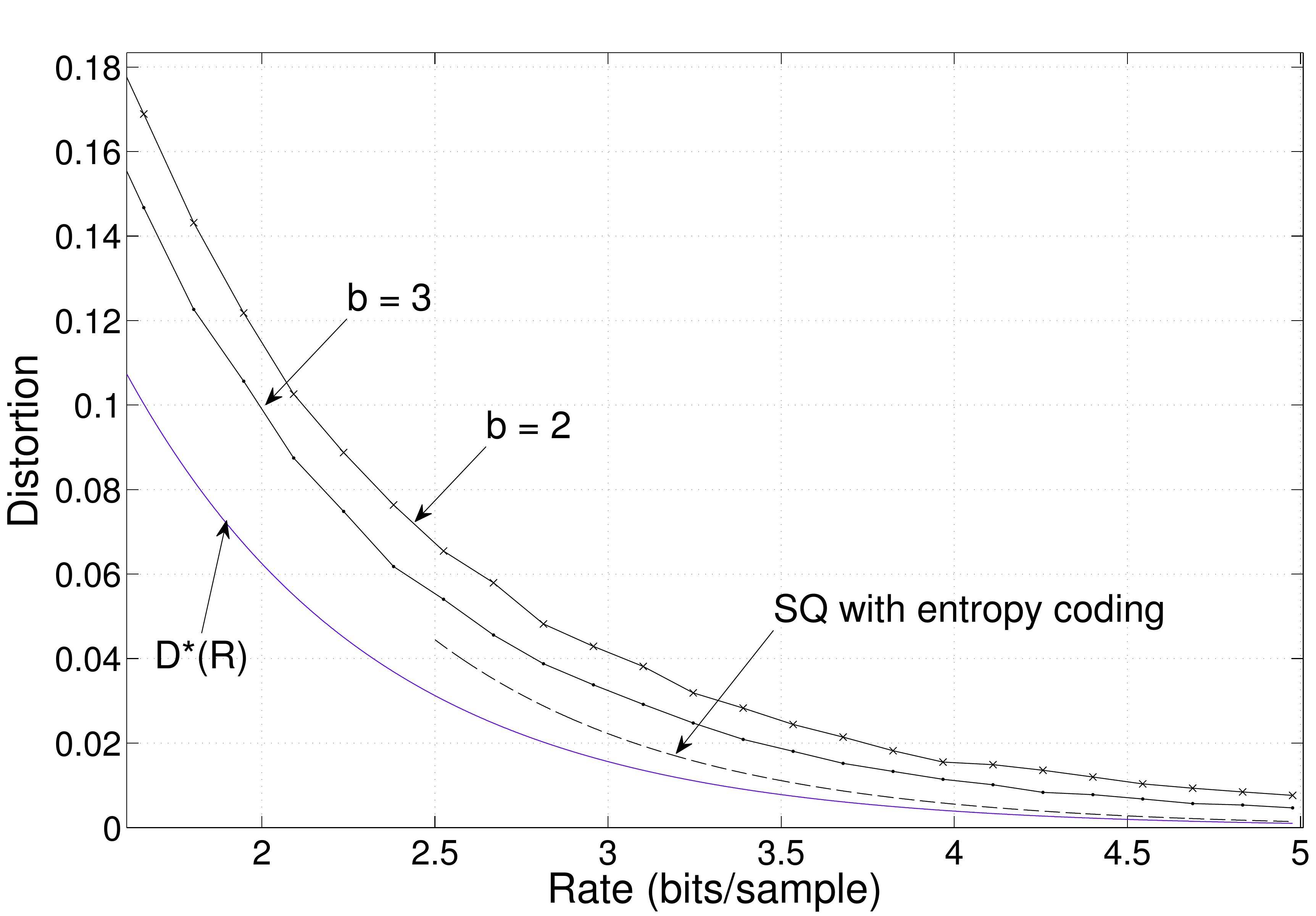}
\caption{\small{Top: Average distortion of the proposed encoder for i.i.d $\mc{N}(0,1)$ source. The design matrix has dimension $n \times ML$ with $M=L^b$. The distortion-rate performance is shown for $b=2$ and $b=3$ along with $D^*(R)=e^{-2R}$.
Bottom: Focusing on the higher rates. The dashed line is the high-rate approximation for the distortion-rate function of an optimal entropy-coded scalar quantizer. }}
\label{fig:sparc_perf}
\end{figure}

\subsection{Simulation Results} \label{subsec:sim}
In this section, we study the performance of the encoder via simulations  on source sequences of unit variance. A brief remark before we proceed.
For the simulations, we use a slightly modified version of the algorithm presented in Section \ref{sec:alg_desc}. In each step $i$, we replace the column selection criterion in \eqref{eq:max_stepi} with
\be
\begin{split}
 m_i  & =  \underset{j: \ (i-1)M < \; j \; \leq iM }{\operatorname{argmin}} \norm{\mbf{R}_{i-1} - c_i \mbf{A}_j }^2 \\ 
 &  = \underset{j: \ (i-1)M < \; j \; \leq iM }{\operatorname{argmax}} 2 c_i \left\langle \mbf{R}_{i-1}, \mbf{A}_j \right\rangle - c_i^2\norm{\mbf{A}_j}^2.
\end{split}
\label{eq:mod_col_sel}
\ee
When $n$ is large, this is almost the same as the original version in \eqref{eq:max_stepi} since  $\norm{\mbf{A}_j}^2 \approx 1$ for all $j$. We found the modified version to have slightly better empirical performance (discussed below), but the original algorithm in Section \ref{sec:alg_desc} is more amenable to theoretical analysis.

The top graph in Fig. \ref{fig:sparc_perf} shows the performance of the proposed encoder on a unit variance i.i.d Gaussian source. The dictionary dimension is $n \times M L$ with $M=L^b$.  The curves show the average distortion at various rates for $b=2$ and $b=3$. The average was obtained from $70$ random trials at each rate. Following convention, rates are plotted in bits rather than nats. The value of $L$ was increased with rate in order to keep the total computational complexity ($\propto nL^{b+1}$) similar across different rates. Recall from \eqref{eq:ml_nR} that the block length is determined by
\[ n = \frac{b L \log L}{R}. \]
For example, for the rates  $1.082, 2.092, 3.102$ and $4.112$ bits/sample, $L$ was chosen to be $46,66,81$ and $97$, respectively.   The corresponding values for the block length are $n=705,573,497,468$ for $b=3$, and $n=470,382,331,312$ for $b=2$. The graph shows the reduction in distortion obtained by increasing $b$ from $2$ to $3$. This reduction comes at the expense of an increase in computational complexity by a factor of $L$.
Simulations were also performed for a unit variance Laplacian source. The resulting distortion-rate curve was virtually identical to Fig. \ref{fig:sparc_perf}, which is consistent with Theorem \ref{thm:rd_feasible}.

As mentioned above, the modified column selection rule given by \eqref{eq:mod_col_sel} has slightly better empirical performance than the original maximum-correlation rule in \eqref{eq:max_stepi}. E.g., for the rates $0.793,1.803$, and $2.957$ bits/sample, the distortions for $b=3$  were $(0.397, 0.123, 0.033)$ with the modified rule, and $(0.406,0.129,0.036)$  with the original rule in \eqref{eq:max_stepi}.
The difference is due to the deviations of the norms of the columns $\mathbf{A}_j$ from $1$. In the parlance of vector quantizer design, one may interpret the modified rule \eqref{eq:mod_col_sel} as taking into account ``gain" in addition to the ``shape" of the source sequence (see, e.g., \cite{HamkZ02}).

 Gish and Pierce \cite{GishPierce68} showed that uniform quantizers with entropy coding are nearly optimal at high rates and that their distortion for a unit variance source is well-approximated by $\frac{\pi e}{6} e^{-2R}$. ($R$ is the entropy of the quantizer in nats.) The bottom graph of Fig. \ref{fig:sparc_perf} zooms in on the higher rates and shows the above high-rate approximation for the distortion of an optimal entropy-coded scalar quantizer (EC-SQ). Recall from \eqref{eq:delta_min} that the distortion gap from $D^*(R)$ is of the order of \footnote{The constants in Theorem \ref{thm:rd_feasible} are not optimized, so the theorem does not give a very precise estimate of the excess distortion in the high-rate, low-distortion regime.}
 \[ \delta_2 \approx \frac{\log \log M}{2 \log M} = \frac{\log b  + \log \log L}{2 b \log L}, \] which is comparable to the optimal $D^*(R)= e^{-2R}$ in the high-rate region. (In fact, $\delta_2$ is larger than $D^*(R)$ at rates greater than $3$ bits for the values of $L$ and $b$ we have used.) This explains the large  {ratio} of the empirical distortion to $D^*(R)$ at higher rates. 

 In summary, the proposed encoder has good empirical performance, especially at low to moderate rates even with modest values of $L$ and $b$. At high rates, there are a few other compression schemes including EC-SQs and the shape-gain quantizer of \cite{HamkZ02} whose empirical rate-distortion performance is close to optimal (see \cite[Table III]{HamkZ02}).  It is shown in  \cite{RVGaussianRD12}  that with minimum-distance encoding, SPARCs attain $D^*(R)$ with the optimal error exponent.   Hence designing computationally feasible SPARC encoders with smaller gap from $D^*(R)$ is a interesting direction for future work.

\section{Proof of Theorem \ref{thm:rd_feasible}} \label{sec:proof}

The essence of the proof is in analyzing the deviation from the typical values of the residual distortion at each step of the encoding algorithm.  In particular, we have to deal with atypicality concerning the source, the design matrix and the maximum  computed in each step of the algorithm.

We introduce some notation to capture the deviations from the typical values. The normalized Euclidean norm of the source is expressed as
\be
\abs{\mbf{S}}^2 = \abs{\mbf{R}_i}^2 = \sigma^2(1+\Delta_0)^2.
\ee
The norm of the residue at stage $i=1, \ldots, L$ is given by
\be
\abs{\mbf{R}_i}^2 = \sigma^2 \left( 1- \frac{2R}{L} \right)^i (1+ \Delta_i)^2.
\label{eq:Deli_def}
\ee
 $\Delta_i \in [-1, \infty)$ measures the deviation of the residual distortion $\abs{\mbf{R}_i}^2$  from its typical value given in \eqref{eq:Ri1_Ri}.

We express the norm of $\mbf{A}_{m_i}$, the column of $\mbf{A}$ chosen in step $i$, as
\be
\abs{\mbf{A}_{m_i}}^2 = 1+ \gamma_i, \quad i=1,\ldots, L.
\label{eq:Ami_dev}
\ee
Recall that the statistics $T^{(i)}_j$ defined  in \eqref{eq:stat_Tij} are i.i.d $\mc{N}(0,1)$ random variables for $j \in \{(i-1)M+1, \ldots, iM \}$.
We write
\be
\begin{split}
\max_{(i-1)M+1 \leq \, j \, \leq iM} \, T^{(i)}_{j} & = \left\langle \mbf{A}_{m_i}, \frac{\mbf{R}_{i-1}}{\norm{\mbf{R}_{i-1}}} \right\rangle \\
& = \sqrt{2 \log M}(1+ \epsilon_i), \quad i=1, \ldots, L.
\end{split}
\ee
The $\epsilon_i$ measure the deviations of the maximum from $\sqrt{2\log M}$ in each step.

Armed with this notation, we have from \eqref{eq:gen_residue}
\be
\begin{split}
& \abs{\mathbf{R}_i}^2   = \abs{\mathbf{R}_{i-1}}^2 + c_i^2 \abs{\mathbf{A}_{m_i}}^2  - \frac{2 c_i \norm{\mathbf{R}_{i-1}}}{n}
\left\langle \mathbf{A}_{m_i}, \frac{\mathbf{R}_{i-1}}{\norm{\mathbf{R}_{i-1}}} \right\rangle\\
&= \sigma^2\left(1 - \frac{2R}{L}\right)^{i-1} (1 + \Delta_{i-1})^2 + c_i^2 (1+ \gamma_i) \\
& \quad - {2c_i \sigma \left(1 - \frac{2R}{L}\right)^{\frac{i-1}{2}} (1+\Delta_{i-1})} \sqrt{\frac{2\log M}{n}} (1+\epsilon_{i}) \\
&= \sigma^2\left(1 - \frac{2R}{L}\right)^{i-1} (1 + \Delta_{i-1})^2   \\
& \quad + \frac{2R \sigma^2}{L} \left(1 - \frac{2R}{L}\right)^{i-1}  (1+ \gamma_i)  \\
& \quad  - \frac{4R\sigma^2}{L} \left(1 - \frac{2R}{L}\right)^{i-1} (1+\Delta_{i-1})(1+\epsilon_i) \\
&= \sigma^2\left(1 - \frac{2R}{L}\right)^{i} \Bigg( (1+\Delta_{i-1})^2  \\
& \qquad  +  \frac{2R/L}{1 - 2R/L}( \Delta_{i-1}^2 + \gamma_i -2\epsilon_i(1+\Delta_{i-1})) \Bigg).
\end{split}
\label{eq:expand_R1}
\ee
From \eqref{eq:expand_R1}, we obtain
\be
\begin{split}
& (1+\Delta_i)^2 \\
&  = (1+\Delta_{i-1})^2 + \frac{2R/L}{1 - 2R/L}( \Delta_{i-1}^2 + \gamma_i -2\epsilon_i(1+\Delta_{i-1}) )
\end{split}
\label{eq:Deli_Deli1}
\ee
for  $i=1, \ldots, L$. The goal is to bound the final distortion  given by
\be \abs{\mbf{R}_L}^2 = \sigma^2 \left( 1 - \frac{2R}{L}\right)^L (1 + \Delta_L)^2 .\ee
We would like to find an upper bound for $(1 + \Delta_L)^2$ that holds under an event whose probability is close to $1$.
Accordingly, define $\mc{A}$ as the event where \emph{all} of the following hold:
\begin{enumerate}
\item $\abs{\Delta_0} < \delta_0$,
\item $\sum_{i=1}^L \frac{\abs{\gamma_i}}{L}< \delta_1$ ,
\item $\sum_{i=1}^L \frac{\abs{\epsilon_i}}{L} < \delta_2$
\end{enumerate}
for $\delta_0, \delta_1, \delta_2$ that satisfy \eqref{eq:del0del1del2}. We upper bound the probability of the event $\mc{A}^c$  using the following  lemmas.

\begin{lem}
For $\delta \in (0,1]$, \[ P\left( \frac{1}{L} \sum_{i=1}^L \abs{\gamma_i}  > \delta \right) < 2 ML \; \exp\left( -n{\delta^2}/{8} \right). \]
\label{lem:gamma_i}
\end{lem}
\begin{IEEEproof}
In Appendix \ref{app:lem_gamma}.
\end{IEEEproof}

\begin{lem} \label{lem:eps_i}
For $\delta>0$, $P\left( \frac{1}{L} \sum_{i=1}^L \abs{\epsilon_i}  > \delta \right) <  \left( \frac{M^{2\delta}}{8 \log M} \right)^{-L}$.
\end{lem}
\begin{IEEEproof}
In Appendix \ref{app:lem_eps}.
\end{IEEEproof}

Using these lemmas, we have
\be
P(\mc{A}^c) < p_0 + p_1 + p_2
\label{eq:PAc}
\ee
where $p_0, p_1, p_2$ are given by \eqref{eq:p0p1p2}.
The remainder of the proof consists of obtaining a bound for  $(1 + \Delta_L)^2$  under the condition that $\mc{A}$ holds. We start with the following lemma.
\begin{lem}
For all sufficiently large $L$, when $\mc{A}$ holds  we have
\be
\Delta_{i} \geq \Delta_0  - \frac{4R}{1-2R/L} \left( \sum_{j=1}^{i} \frac{\abs{\gamma_j} + \abs{\epsilon_j}}{L}\right), \quad i=1, \ldots,L.
\label{eq:Deli_ind}
\ee
In particular, $\Delta_i > -\frac{1}{2}, \quad i=1, \ldots ,L$
\label{lem:Deli_LB}
\end{lem}
\begin{IEEEproof}
We first show that $\Delta_i > -\frac{1}{2}$ follows from \eqref{eq:Deli_ind}. Indeed, \eqref{eq:Deli_ind} implies that
\be
\begin{split}
\Delta_{i} & \geq \Delta_0  - \frac{4R}{1-2R/L} \left( \sum_{j=1}^{i} \frac{\abs{\gamma_j} + \abs{\epsilon_j}}{L}\right) \\
& \stackrel{(a)}{>} - \delta_0 - {5R}\left(\delta_1  + \delta_2 \right) \stackrel{(b)}{>}  -\frac{1}{2} \label{eq:final_deli}
\end{split}
\ee
where $(a)$ is obtained from the conditions of $\mc{A}$ while $(b)$ holds due to \eqref{eq:del0del1del2}.

We now prove \eqref{eq:Deli_ind} by induction. The statement trivially holds for $i=0$. Towards induction, assume \eqref{eq:Deli_ind} holds for $i-1$ for some $i \in \{1, \ldots, L\}$. From \eqref{eq:Deli_Deli1}, we obtain
\be
\begin{split}
& (1+\Delta_i)^2 \\
&= (1+\Delta_{i-1})^2 + \frac{2R/L}{1 - 2R/L}( \Delta_{i-1}^2 +  \gamma_i - 2\epsilon_i(1+\Delta_{i-1}) ) \\
& \geq (1+\Delta_{i-1})^2  -  \frac{2R/L}{1 - 2R/L}(\abs{\gamma_i} + 2\abs{\epsilon_i}(1+\Delta_{i-1}) ).
\end{split}
\label{eq:st1}
\ee
For $L$ large enough, the right side above is positive and we therefore have
{\small{
\be
\begin{split}
& (1 + \Delta_i)  \\
& \geq (1+ \Delta_{i-1})
\left[ 1 - \frac{2R/L}{1 - 2R/L} \left[ \frac{\abs{\gamma_i}}{(1+\Delta_{i-1})^2} + \frac{2\abs{\epsilon_i}}{1+\Delta_{i-1}} \right)  \right]^{\frac{1}{2}} \\
& \geq (1+\Delta_{i-1})\left[ 1 - \frac{2R/L}{1 - 2R/L} \left( \frac{\abs{\gamma_i}}{(1+\Delta_{i-1})^2} + \frac{2\abs{\epsilon_i}}{1 +\Delta_{i-1}} \right) \right] \\
& =  1 + \Delta_{i-1} - \frac{2R/L}{1 - 2R/L} \left( \frac{\abs{\gamma_i}}{(1+\Delta_{i-1})} +  {2\abs{\epsilon_i}} \right)
\end{split}
\label{eq:st2}
\ee
}}
where the second  inequality holds since $\sqrt{1-x} \geq 1-x$  for $x\in(0,1)$. \eqref{eq:st2} implies that
\be
\begin{split}
\Delta_i & \geq \Delta_{i-1} - \frac{2R/L}{1-2R/L} \left( \frac{\abs{\gamma_i}}{(1+\Delta_{i-1})} +  {2\abs{\epsilon_{i}}} \right) \\
& \stackrel{(a)}{\geq}  \Delta_{i-1} - \frac{2R/L}{1-2R/L} \left( 2{\abs{\gamma_i}} +  {2\abs{\epsilon_{i}}} \right) \\
& \stackrel{(b)}{\geq}  \Delta_0  - \frac{4R}{1-2R/L} \left( \sum_{j=1}^{i-1} \frac{\abs{\gamma_j} + \abs{\epsilon_j}}{L}\right) \\
& \quad  - \frac{4R/L}{1-2R/L}( \abs{\gamma_i} + \abs{\epsilon_i}).
\end{split}
\label{eq:Del1_Del0_rel}
\ee
In the chain above, $(a)$ holds because $\Delta_{i-1} > \frac{1}{2}$, a consequence of the induction hypothesis as shown in \eqref{eq:final_deli}. $(b)$ is obtained by using the induction hypothesis for $\Delta_{i-1}$. The proof of the lemma is complete.
\end{IEEEproof}
\begin{lem}
When $\mc{A}$ is true and $L$ is large enough that Lemma \ref{lem:Deli_LB} holds,
\be  \abs{\Delta_i} \leq \abs{\Delta_0} w^i + \frac{4R/L}{1 - 2R/L} \sum_{j=1}^i w^{i-j} (\abs{\gamma_j} + \abs{\epsilon_j}) \ee
for $i=1,\ldots, L$, where $w= \left( 1 + \frac{R/L}{1-2R/L}\right)$.
\label{lem:Ri_UB}
\end{lem}
\begin{IEEEproof}
We prove the lemma by induction. For $i=1$, we have from \eqref{eq:Deli_Deli1}
\be
\begin{split}
& (1+\Delta_1)^2\\
& = (1+\Delta_0)^2 + \frac{\tfrac{2R}{L}}{1 - \tfrac{2R}{L}}(\Delta_0^2 + \gamma_1 -2\epsilon_1(1+\Delta_0)) \\
& \leq 1 + \Delta_0^2 + 2\abs{\Delta_0} +  \frac{\tfrac{2R}{L}}{1 - \tfrac{2R}{L}}( \Delta_0^2 + \abs{\gamma_1} + 2\abs{\epsilon_1}(1 + \abs{\Delta_0}) )\\
& = (1 + \abs{\Delta_0})^2
\left[  1 +   \frac{\tfrac{2R}{L}}{1 - \tfrac{2R}{L}}\left( \frac{\Delta_0^2}{(1 +\abs{\Delta_0})^2} \right. \right. \\
& \hspace{1in} \left.  + \frac{\abs{\gamma_1}}{(1+\abs{\Delta_0})^2} + \frac{2\abs{\epsilon_1}}{(1 + \abs{\Delta_0})} \right)   \Bigg].
\end{split}
\label{eq:ub_step1}
\ee
Therefore,
\be
\begin{split}
& 1 + \Delta_1  \leq (1 + \abs{\Delta_0})\Bigg[  1 +   \frac{2R/L}{1 - 2R/L}\left( \frac{\Delta_0^2}{(1 +\abs{\Delta_0})^2} \right. \\
& \hspace{1.2in} \left. +  \, \frac{\abs{\gamma_1}}{(1+\abs{\Delta_0})^2} + \frac{2\abs{\epsilon_1}}{(1 + \abs{\Delta_0})} \right)   \Bigg]^{\frac{1}{2}} \\
& \leq (1 + \abs{\Delta_0})\Bigg[  1 +   \frac{R/L}{1 - 2R/L}\left( \frac{\Delta_0^2}{(1 +\abs{\Delta_0})^2} \right. \\
& \hspace{1.2in} \left. + \,  \frac{\abs{\gamma_1}}{(1+\abs{\Delta_0})^2} + \frac{2\abs{\epsilon_1}}{(1 + \abs{\Delta_0})} \right)   \Bigg]
\end{split}
\label{eq:ub_step2}
\ee
where we have used the inequality $\sqrt{1+x} \leq 1 + \frac{x}{2}$ for $x>0$. We therefore have
\be
\begin{split}
\Delta_1 & \leq \abs{\Delta_0} + \frac{R/L}{1 - 2R/L}\left( \frac{\Delta_0^2}{(1 +\abs{\Delta_0})} + \frac{\abs{\gamma_1}}{(1+\abs{\Delta_0})} + {2\abs{\epsilon_1}} \right) \\
& \stackrel{(a)}{\leq}  \abs{\Delta_0} + \frac{R/L}{1 - 2R/L} ( \abs{\Delta_0} + \abs{\gamma_1} + 2\abs{\epsilon_1}) \\
& \leq \abs{\Delta_0}\left( 1 +  \frac{R/L}{1 - 2R/L} \right) + \frac{2R/L}{1 - 2R/L}(\abs{\gamma_1} +  \abs{\epsilon_1}).
\end{split}
\label{eq:Del1UB}
\ee
In \eqref{eq:Del1UB}, we used $\abs{\Delta_0}/(1 + \abs{\Delta_0}) < 1$ to obtain $(a)$. From Lemma \ref{lem:Deli_LB}, we have
\be
\begin{split}
\Delta_{1} & \geq \Delta_0  - \frac{4R/L}{1-2R/L} \left(\abs{\gamma_1} + \abs{\epsilon_1}\right) \\
&  \geq  - \abs{\Delta_0}  - \frac{4R/L}{1-2R/L} \left(\abs{\gamma_1} + \abs{\epsilon_1}\right).
\end{split}
\label{eq:Del1LB}
\ee
Combining \eqref{eq:Del1UB} and \eqref{eq:Del1LB}, we obtain
\be
\abs{\Delta_1} \leq \abs{\Delta_0}\left( 1 +  \frac{R/L}{1 - 2R/L} \right) + \frac{4R/L}{1 - 2R/L}(\abs{\gamma_1} + \abs{\epsilon_1}).
\label{eq:AbsDel1UB}
\ee
This completes the proof for $i=1$. Towards induction, assume that the lemma holds for $i-1$. From \eqref{eq:Deli_Deli1}, we obtain
\be
\begin{split}
(1+\Delta_i)^2  &  \leq 1   + \Delta_{i-1}^2 + 2\abs{\Delta_{i-1}}  \\
& \  +  \frac{2R/L}{1 - 2R/L}( \Delta_{i-1}^2 + \abs{\gamma_i} + 2\abs{\epsilon_i}(1 + \abs{\Delta_{i-1}}) ).
\end{split}
\ee
Using arguments identical to those in \eqref{eq:ub_step1}--\eqref{eq:Del1UB}, we get
\be
\Delta_i \leq \abs{\Delta_{i-1}}\left( 1 +  \frac{R/L}{1 - 2R/L} \right) + \frac{2R/L}{1 - 2R/L}(\abs{\gamma_i} + \abs{\epsilon_i}).
\label{eq:DeliUB}
\ee
From the proof of Lemma \ref{lem:Deli_LB} (see \eqref{eq:Del1_Del0_rel}), we have
\be
\begin{split}
\Delta_{i} & \geq  \Delta_{i-1} - \frac{4R/L}{1-2R/L} \left( {\abs{\gamma_i}} +  {\abs{\epsilon_{i}}} \right) \\
&  \geq -\abs{\Delta_{i-1}} - \frac{4R/L}{1-2R/L} \left( {\abs{\gamma_i}} +  {\abs{\epsilon_{i}}} \right).
\end{split}
\label{eq:DeliLB}
\ee
Combining \eqref{eq:DeliUB} and \eqref{eq:DeliLB}, we obtain
\be
\abs{\Delta_i} \leq \abs{\Delta_{i-1}}\left( 1 +  \frac{R/L}{1 - 2R/L} \right) + \frac{4R/L}{1 - 2R/L}(\abs{\gamma_i} + \abs{\epsilon_i}).
\label{eq:AbsDeliUB}
\ee
Using the induction hypothesis to bound $\abs{\Delta_{i-1}}$ in \eqref{eq:AbsDeliUB}, we obtain
\ben
\begin{split}
\abs{\Delta_i}  \leq & \left(\abs{\Delta_0} w^{i-1}  +  \frac{4R/L}{1 - 2R/L} \sum_{j=1}^{i-1} w^{i-1-j} (\abs{\gamma_j} + \abs{\epsilon_j}) \right)  \\
& \cdot \left(1 +  \frac{R/L}{1 - 2R/L} \right)  +   \frac{4R/L}{1 - 2R/L}(\abs{\gamma_i} + \abs{\epsilon_i}) \\
& = \abs{\Delta_0} w^i + \frac{4R/L}{1 - 2R/L} \sum_{j=1}^i w^{i-j} (\abs{\gamma_j} + \abs{\epsilon_j}).
\end{split}
\een
\end{IEEEproof}
Lemma \ref{lem:Ri_UB} implies that when $\mc{A}$ holds and $L$ is sufficiently large,
\be
\begin{split}
\abs{\Delta_L} & \leq  \abs{\Delta_0} w^L + \frac{4R/L}{1 - 2R/L} \sum_{j=1}^L w^{L-j} (\abs{\gamma_j} + \abs{\epsilon_j}) \\
& \leq w^L \left[ \abs{\Delta_0} + \frac{4R}{(1-2R/L)w}  \left(\sum_{j=1}^L \frac{\abs{\gamma_j}}{L}  +  \sum_{j=1}^L  \frac{\abs{\epsilon_j}}{L} \right)  \right] \\
& \stackrel{(a)}{\leq}  w^L \left[  {\delta_0} + \frac{4R}{(1-R/L)} (\delta_1 + \delta_2) \right] \\
& \stackrel{(b)}{\leq} \exp\left(\frac{R}{1 - 2R/L}\right) \left[  \delta_0 + \frac{4R}{(1-R/L)} (\delta_1 + \delta_2)  \right] \\
& \leq e^R \left(\delta_0 + 5R(\delta_1 + \delta_2) \right) \quad \text{ for  large enough } L.
\end{split}
\label{eq:Del_L_UB}
\ee
$(a)$ is true because $\mc{A}$ holds and $(b)$ is obtained by applying the inequality $1 + x \leq e^x$ with $x=\frac{R/L}{1 - 2R/L}$.

Hence when $\mc{A}$ holds and $L$ is sufficiently large, the distortion can be bounded as
\be
\begin{split}
\abs{R_L}^2  = \sigma^2 e^{-2R}(1 + \Delta_L)^2 & \leq \sigma^2 e^{-2R} (1 + \abs{\Delta_L})^2 \\
&  \stackrel{(c)}{\leq} \sigma^2 e^{-2R} (1 + e^R {\Delta})^2 
\end{split}
\label{eq:finalRL}
\ee
where $(c)$ follows from \eqref{eq:Del_L_UB} by defining $\Delta = \delta_0 + 5R(\delta_1 + \delta_2)$. Combining \eqref{eq:finalRL} with \eqref{eq:PAc}
completes the proof of the theorem.

\section{Discussion} \label{sec:conc}

We have studied a new ensemble of codes for lossy compression where the codewords are structured linear combinations of elements of a design matrix. The size of the design matrix is a low-order polynomial in the block length, as a result of which the storage complexity is much lower than that of the random i.i.d codebook. We proposed a  successive-approximation  encoder with computational complexity growing polynomially in the block-length.  For any ergodic source with known variance, the encoder was shown to attain $D^*(R)$, the optimal distortion-rate function of an i.i.d Gaussian source with the same variance. Further, if the second moment of the source satisfies a large deviations property,  the probability of excess distortion (for any fixed distortion-level greater than $D^*(R)$) decays exponentially with the block length.

The encoding algorithm may be interpreted as successively refining the source over an asymptotically large number of stages with asymptotically small rate in each stage. We emphasize that the successive refinement interpretation is unique to this particular algorithm, and is not an inherent property of the sparse regression codebook. The section coefficients $c_i$ were chosen to optimize the encoding algorithm. The coefficients allocate `power' across sections of the design matrix and they are chosen depending on the encoder.   For example, the optimal (minimum-distance) encoder  analyzed in \cite{RVGaussianRD12}  has equal-valued section coefficients.

 For the proposed encoder, the gap from $D^*(R)$ as a function of design matrix dimension is $O( \log \log M /\log M)$, as given in \eqref{eq:delta_min}. An important direction for future work is designing computationally-efficient encoders for SPARCs with faster convergence to  $D^*(R)$ with the dimension (or block length).  The results of \cite{IngberKochman,KostinaV12} show that the optimal  gap from $D^*(R)$  (among all codes) is $\Theta({1}/{\sqrt{n}})$. The fact that SPARCs achieve the optimal error-exponent with minimum-distance encoding \cite{RVGaussianRD12} suggests that it is possible to design encoders with faster convergence to $D^*(R)$ at the expense of slightly higher computational complexity.  A simple way to improve  on successive refinement encoder is the following: after the algorithm terminates, one may  perform  column swaps within sections  in order to  improve the final distortion.  Another idea is to  make the encoder less `greedy', i.e., search across multiple sections instead of sequentially picking one column at a time. Techniques such as  $\ell_1$-norm based convex optimization \cite{CandesTaoLP, DonohoCS,WainLasso09} and approximate message passing \cite{bayatiM11} which have been successful for sparse signal recovery may also prove useful. Another approach  to improve the high-rate distortion performance is to construct a few sections of the design matrix in a structured way so as to optimize the shapes of the Voronoi cells.

 Another  direction for further investigation is exploring design matrices with smaller storage complexity. For example, a SPARC defined by a design matrix with i.i.d $\pm 1$ entries was found to have empirical distortion-rate performance very similar to the Gaussian design matrix. Since binary entries imply a much reduced storage requirement compared to Gaussian entries, establishing theoretical performance bounds for the $\pm 1$ design matrix is an interesting open problem. For communication over AWGN channels, the performance of a binary SPARC codebook with minimum-distance encoding was recently analyzed by Takeishi et al \cite{TakeishiKT13}.

 The results of this paper  together with those in \cite{AntonyFast} show that SPARCs with computationally efficient encoders and decoders can be used for both lossy compression and communication, at rates approaching the Shannon-theoretic limits. Further, \cite{RVAller12} demonstrates how source and channel coding SPARCs can be nested to implement binning and superposition, which are key ingredients of coding schemes for multi-terminal source and channel coding problems. Sparse regression codes therefore offer a promising framework to develop fast, rate-optimal codes for a variety of models in network information theory.

\appendices
\section{Proof of Lemma  \ref{lem:inner_prod}} \label{app:inner_prod}

The joint density of $T_1,\ldots, T_N$ can be expressed as 
\be \begin{split} 
& f_{T_1, \ldots, T_N}(t_1,\ldots, t_N) \\
& = \int_{\mbf{r}} f_{T_1, \ldots, T_N| \mbf{R}}(t_1,\ldots, t_N \, | \, \mbf{R} = \mbf{r}) \, d \nu(\mbf{r}) \label{eq:fcondT}
\end{split} 
\ee
where $\nu$ is the distribution of $\mbf{R}$, and $f_{T_1, \ldots, T_N| \mbf{R}}$ is the joint density of $T_1, \ldots, T_N$ conditioned on $\mbf{R}$.
Conditioned on
\[ \mbf{R} = \mbf{r} = (r_1,\ldots, r_n), \]
we have for $j=1, \ldots, N$
\be
T_{j}  = \left\langle  \mbf{A}_j, \mbf{r} \right\rangle = r_1 A_{j1} + r_2 A_{j2} \ldots + r_n A_{jn}
\ee
Note that $\{A_{j1}, \ldots, A_{jn}\}$ for $1 \leq j \leq N$ are a collection of $Nn$  random variables that are i.i.d $\mc{N}(0,1)$, and $r_1,\ldots, r_n$ are constants such that $\sum_i r_i^2=1$. Hence conditioned on \emph{any} realization $\mathbf{R} =\mbf{r}$,  $T_{j}, \, 1 \leq j \leq N$  are mutually independent $\mc{N}(0,1)$ random variables. Therefore, the conditional joint density in \eqref{eq:fcondT} becomes
\be 
f_{T_1, \ldots, T_N| \mbf{R}}(t_1,\ldots, t_N \, | \, \mbf{R} = \mbf{r}) = \prod_{j=1}^N \phi(t_j), \quad \forall \mbf{r}
\label{eq:cond_ind_gauss}
\ee
where $\phi(.)$ denotes the density of a $\mc{N}(0,1)$ random variable. Using \eqref{eq:cond_ind_gauss} in \eqref{eq:fcondT}, we obtain
\be
f_{T_1, \ldots, T_N}(t_1,\ldots, t_N) = \int_{\mbf{r}} \, \prod_{j=1}^N \phi(t_j) \, d \nu(\mbf{r}) = \prod_{j=1}^N \phi(t_j).
\ee

\section{Proof of Lemma \ref{lem:gamma_i}} \label{app:lem_gamma}
Recall from \eqref{eq:Ami_dev} that
\[ \gamma_i = \abs{\mbf{A}_{m_i}}^2 - 1, \quad i=1,\ldots,L. \]
We have
\be
\begin{split}
& P\left( \frac{1}{L} \sum_{i=1}^L \abs{\gamma_i}  > \delta \right)   \leq  P\left( \cup_{i=1}^L \{ \abs{\gamma_i}  > \delta \} \right)  \\
& \leq \sum_{i=1}^L P\left( \abs{\gamma_i}  > \delta \right) \\
&= \sum_{i=1}^L P\left( \left\{ \abs{\mbf{A}_{m_i}}^2   >  1+ \delta \right\} \; \cup \; \left\{ \abs{\mbf{A}_{m_i}}^2   <  1 - \delta \right\} \right).
\end{split}
\ee
The right-side above can be bounded as
\be
\begin{split}
&\sum_{i=1}^L P\left( \left\{ \abs{\mbf{A}_{m_i}}^2   >  1+ \delta \right\} \; \cup \; \left\{ \abs{\mbf{A}_{m_i}}^2   <  1 - \delta \right\} \right)  \\
&\stackrel{(a)}{\leq} \sum_{i=1}^L P\left( \cup_{j = (i-1)M+1}^{iM} \ \left\{ \abs{\mbf{A}_{j}}^2   >  1+ \delta \right\} \right. \\
& \hspace{1.5 in} \cup  \left\{ \abs{\mbf{A}_{j}}^2   <  1 - \delta \right\} \Big)\\
& \stackrel{(b)}{\leq} \sum_{i=1}^L \sum_{j = (i-1)M+1}^{iM} \hspace{-3pt}\left( P\left( \abs{\mbf{A}_{j}}^2   >  1+ \delta \right)
+ P \left( \abs{\mbf{A}_{j}}^2   <  1 - \delta \right) \right) \\
&= ML \Big( P\left( \abs{\mbf{A}_{j}}^2   >  1+ \delta \right) + P \left( \abs{\mbf{A}_{j}}^2   <  1 - \delta \right) \Big).
\end{split}
\label{eq:gamma_dev0}
\ee
$(a)$ follows from the observation that $m_i \in \{(i-1)M+1, \ldots, iM \}$ , i.e., $\mbf{A}_{m_i}$ is one of the columns on Section $i$ of $\mbf{A}$.
$(b)$ is due to the union bound.

Using a Chernoff bound for $ P\left( \abs{\mbf{A}_{m_i}}^2   >  1+ \delta \right)$, we have
\be
\begin{split}
& P\left( \abs{\mbf{A}_{j}}^2  >  1+ \delta \right)  =  P( \norm{\mbf{A}_{j}}^2  >  n(1+ \delta)) \\
 & \quad \leq \exp( -t n (1+ \delta)) \; \expec[\exp(t \norm{\mbf{A}_{j}}^2 )], \quad t>0 \\
 &\quad  = \exp( -t n (1+ \delta))\; (1-2t)^{-n/2}.
\end{split}
\ee
The last line is obtained by using the  moment generating function of $\norm{\mbf{A}_{j}}^2$, a $\chi^2_n$ random variable. Using $t= \frac{\delta}{2(1+\delta)}$, we get
\be
P\left( \abs{\mbf{A}_{j}}^2  >  1+ \delta \right) \leq \exp\left(  \frac{-n\delta}{2}\right) (1+\delta)^{n/2} \leq \exp\left( \frac{-n\delta^2}{8} \right)
\label{eq:pos_del}
\ee
where the second inequality above is obtained using the bound $\ln(1 + \delta) \leq \delta - \frac{\delta^2}{4}$ for $\delta \in [0,1]$.

Similarly,
\be
\begin{split}
 & P\left( \abs{\mbf{A}_{j}}^2  <  1 - \delta \right)  =  P( \norm{\mbf{A}_{j}}^2  <  n(1 - \delta) ) \\
 & \quad \leq \exp( t n (1 - \delta)) \; \expec[\exp( - t \norm{\mbf{A}_{j}}^2 )], \quad t>0 \\
 & \quad  = \exp( t n (1 - \delta))\; (1 + 2t)^{-n/2}.
\end{split}
\ee
Using $t= \frac{\delta}{2(1 - \delta)}$, we get
\be
P\left( \abs{\mbf{A}_{j}}^2  <  1 - \delta \right) \leq \exp\left(  \frac{n\delta}{2}\right) (1 - \delta)^{n/2} \leq \exp\left(  \frac{-n\delta^2}{4} \right)
\label{eq:neg_del}
\ee
where we have used $\log(1 - \delta) \leq -\delta - \frac{\delta^2}{2}$ for $\delta \in [0,1]$.
Substituting \eqref{eq:pos_del} and \eqref{eq:neg_del} in \eqref{eq:gamma_dev0} completes the proof.

\section{Proof of Lemma \ref{lem:eps_i}} \label{app:lem_eps}
For a random variable $X$, let $f_X$ and $F_X$ denote the density and distribution functions, respectively.
Recall from \eqref{eq:stat_Tij} that for $i \in \{1, \ldots, L\}$ and  $j \in \{ (i-1)M+1, \ldots, iM \}$, the statistic
\be T^{(i)}_{j} = 
\left\langle \mbf{A}_j, \frac{\mbf{R}_{i-1}}{\norm{\mbf{R}_{i-1}}} \right\rangle. \label{eq:stat_Tij2}
\ee
Define for $i=1, \ldots, L$,
\be
Z_i = \max_{(i-1)M+1 \, \leq j \, \leq iM} \ T^{(i)}_{j} = \sqrt{2 \log M}(1 + \e_i).
\label{eq:Zi_def}
\ee
We first show that the $Z_i$'s in \eqref{eq:Zi_def} are i.i.d and thus
\be
\e_i = \frac{Z_i}{\sqrt{2 \log M}} -1, \quad i=1, \ldots, L
\label{eq:ei_def}
\ee
are i.i.d random variables. For brevity, we denote the collection $\{T^{(i)}_j, \ (i-1)M+1 \leq j \leq iM \}$ by $T^{(i)}$ for $i=1, \ldots, L$.
Consider the conditional joint distribution function $F_{T^{(i)} \mid T^{(i-1)}, \ldots,  T^{(1)}, \mathbf{R}_0}$. We have
\be
\begin{split}
F_{T^{(i)} \mid T^{(i-1)}, \ldots,  T^{(1)}, \mathbf{R}_0}  \stackrel{(a)}{=} F_{T^{(i)} \mid \mathbf{R}_{i-1}} \
\ \stackrel{(b)}{=} \ \prod_{j=(i-1)M+1}^{iM} F_{T^{(i)}_j}.
\end{split}
\label{eq:Ti_indep}
\ee
The   equality  $(a)$ is obtained by using the following two observations about $T^{(i)}_{j} $ in  \eqref{eq:stat_Tij2}: 1) each column $\mbf{A}_j$ in the  $i$th section of $\mbf{A}$ is independent of
$\{T^{(i-1)}, \ldots,  T^{(1)}, \mathbf{R}_0=\mathbf{S}\}$ because the latter are functions of the source sequence and the columns in the first $i-1$ sections of $\mbf{A}$; 2)  for $i\geq 1$, $\mbf{R}_{i-1}$ is a function of $\{T^{(i-1)}, \ldots,  T^{(1)}, \mathbf{R}_0=\mathbf{S}\}$.
In \eqref{eq:Ti_indep}, $(b)$ follows from  Lemma \ref{lem:inner_prod}; recall that for each $i$, conditioned on $\mathbf{R}_{i-1} = \mathbf{r}$,  the random variables
$\{ T^{(i)}_j,  \ (i-1)M+1 \leq j \leq iM \}$ are i.i.d $\mc{N}(0,1)$.
Therefore
\[ F_{ \mathbf{R}_0, T^{(1)},  \ldots, T^{(L)}} = F_{\mathbf{R}_0} \prod_{i=1}^L \prod_{j=(i-1)M+1}^{iM} F_{T^{(i)}_j}  \]
where $\{T^{(i)}_j\} \sim $ i.i.d $\mc{N}(0,1)$ $\forall i,j$. Consequently,  the $\{ Z_i, \ 1\leq i \leq L\}$ are i.i.d random variables.

Using a Chernoff bound, we have
\be P\left(\sum_{i=1}^{L}\frac{\abs{\epsilon_i}}{L} > \delta\right) \leq  \left( \expec[\exp(t\abs{\epsilon_1})] \; {\exp(-t\delta)}\right)^L, \quad \forall t>0. \label{eq:chernoff0} \ee
We choose $t= 2\log M$ and compute the bound. We have
\be
\begin{split}
&\expec[\exp(t\abs{\epsilon_1})]\\ 
& = \, \int_{0}^{\infty} \exp(tx) f_{\epsilon_1}(x)dx + \int_{-\infty}^{0}\exp(-tx)f_{\epsilon_1}(x)dx.
\end{split}
\label{eq:int_split}
\ee
The first integral can be bounded as follows.
\be
\begin{split}
& \int_{0}^{\infty}\exp(tx)f_{\epsilon_1}(x)dx  \leq \int_{-\infty}^{\infty}\exp(tx) f_{\epsilon_1}(x)dx \\
& \qquad  {=}  \ \expec\left[ \exp(t \epsilon_1) \right]
 {=} \ \frac{1}{M^2} \expec[\exp(\sqrt{2 \log M} \; Z_1)]%
\end{split}
\label{eq:first_I}
\ee
where the last equality is obtained from \eqref{eq:ei_def} and $t=2\log M$.
Since $Z_1$ is the maximum of $\mc{N}(0,1)$ i.i.d random variables $T^{(1)}_j, \; 1\leq j\leq M$, we have
\be
\begin{split}
\expec\left[ \exp( \sqrt{2 \log M} \; Z_1 ) \right] & =
\expec\left[ \max_{j} \; \exp( \sqrt{2 \log M} \;  T^{(1)}_j ) \right] \\
&\leq \expec\left[ \sum_{j}  \exp( \sqrt{2 \log M} \; T^{(1)}_j ) \right] \\
&= M \expec\left[ \exp( \sqrt{2 \log M} \; T^{(1)}_1 ) \right]\\
& \stackrel{(a)}{=} M  \exp\left( \frac{(\sqrt{2 \log M})^2}{2} \right) = M^2
\end{split}
\label{eq:mgf_bound}
\ee
where $(a)$ is obtained by  evaluating the moment-generating function of a $\mc{N}(0,1)$ random variable at $\sqrt{2 \log M}$. Using
\eqref{eq:mgf_bound} in \eqref{eq:first_I}, we obtain
\be \int_{0}^{\infty}\exp(tx)f_{\epsilon_1}(x)dx \leq 1. \label{eq:final_boundI0} \ee

The second integral in \eqref{eq:int_split} can be written as
\be 
\begin{split} 
& \int_{-\infty}^{0}\exp(-tx)f_{\epsilon_1}(x)dx \\
& = \int_{-\infty}^{0}\exp(-tx)f_{Z_1}(\sqrt{2 \log M}\; (x+1)) \, \sqrt{2 \log M} dx  
\end{split} 
\ee
where we have used \eqref{eq:ei_def} to express $f_{\epsilon_1}$ in terms of $f_{Z_1}$. Using the change of variable $z=\sqrt{2\log{M}}(x + 1)$, we have
\be
\begin{split}
& \int_{-\infty}^{0}\exp(-tx)f_{\epsilon_1}(x)dx \\
& = \int_{-\infty}^{\sqrt{2\log{M}}}\exp\left(-t\left(\dfrac{z}{\sqrt{2\log{M}}} - 1\right)\right)f_{Z_1}(z)dz \\
& =I_1 + I_2 + I_3
\end{split}
\ee
where
\begin{align}
I_1 &= \int_{-\infty}^{0}\exp\left(-t\left(\dfrac{z}{\sqrt{2\log{M}}} - 1\right)\right)f_{Z_1}(z)dz, \\
I_2 &=  \int_{0}^{\sqrt{2\log{M}}-1}\exp\left(-t\left(\dfrac{z}{\sqrt{2\log{M}}} - 1\right)\right)f_{Z_1}(z)dz, \\
I_3 &= \int_{\sqrt{2\log{M}}-1}^{\sqrt{2\log{M}}}\exp\left(-t\left(\dfrac{z}{\sqrt{2\log{M}}} - 1\right)\right)f_{Z_1}(z)dz.
\end{align}
We evaluate each of these integrals below.  Since $Z_1$ is the maximum of $M$ standard Gaussians, its distribution function and density are given by
\[ F_{Z_1}(z) = (\Phi(z))^M, \quad f_{Z_1}(z) = M \phi(z) (\Phi(z))^{M-1} \]
where $\Phi$ and $\phi$ denote the standard Gaussian distribution function and density, respectively.

$I_1$ can then be written as
\be
\begin{split}
I_1 & =\int_{-\infty}^{0}  \exp\left(-t\left(\dfrac{z}{\sqrt{2\log{M}}} - 1\right)\right) M (\Phi(z))^{M-1} \phi(z)dz \\
&\stackrel{(a)}{=} \frac{M}{2^{M-1}}  \int_{-\infty}^{0}  \exp\left(-t\dfrac{z}{\sqrt{2\log{M}}}\right) \exp(t) \phi(z)dz \\
&\stackrel{(b)}{=} \frac{M^3}{2^{M-1}} \int_{-\infty}^{0} \exp\left(-\sqrt{2\log{M}}z\right) \phi(z)dz \\
&{=} \frac{M^3}{2^{M-1}} \int_{0}^{\infty} \exp\left(\sqrt{2\log{M}}z\right) \phi(z)dz \\
&\leq \frac{M^3}{2^{M-1}}  \int_{-\infty}^{\infty} \exp\left(\sqrt{2\log{M}}z\right) \phi(z)dz \\
&\stackrel{(c)}{=} \frac{M^4}{2^{M-1}}.
\end{split}
\label{eq:I1_bound}
\ee
In the above, $(a)$ is true because $\Phi(z) \leq \dfrac{1}{2}$ for $z \leq 0$, $(b)$ is obtained by substituting $t = 2 \log M$, and $(c)$ is obtained by evaluating the moment generating function of a standard Gaussian at $\sqrt{2\log{M}}$.

Next,
\be
\begin{split}
I_2 & = \int_{0}^{\sqrt{2\log{M}} - 1} \exp\left(-t\left(\dfrac{z}{\sqrt{2\log{M}}} - 1\right)\right)\\
& \hspace{1in} \cdot M (\Phi(z))^{M-1} \phi(z)dz \\
& \leq M^3 \left( \max_{z \in [0,\sqrt{2\log M} -1]} \; \exp\left(-\sqrt{2\log{M}}z\right) (\Phi(z))^{M-1}\right)\\
& \quad \cdot  \int_{0}^{\sqrt{2\log{M}} - 1} \phi(x) dx.
\end{split}
\label{eq:I2_begin}
\ee
Let
\[ g(z) = \exp\left(-\sqrt{2\log{M}}z\right) (\Phi(z))^{M-1}. \]
It can be verified that $g(z)$ is an increasing function in $z \in [0,\sqrt{2\log M} -1]$ for large enough $M$ ($M>6$ is sufficient). Therefore the maximum is attained at  $\sqrt{2\log M} -1$ and \eqref{eq:I2_begin} becomes
\be
\begin{split}
& I_2 \leq M^3 g(\sqrt{2\log M} -1) \\
& = M \exp(\sqrt{2\log{M}}) (\Phi(\sqrt{2 \log M} -1))^{M-1}.
\end{split}
\label{eq:I2_UB}
\ee

\emph{\textbf{Claim} $\mathbf{1}$}: $I_2 \to 0$ as $M \to \infty$.
\begin{IEEEproof}
Using the bound
\be \Phi(x) \leq \left(1 - \dfrac{x}{1 + x^2}\dfrac{\exp(-\frac{x^2}{2})}{\sqrt{2\pi}}\right), \label{eq:Phi_bound} \ee
we have
\be
\begin{split}
& \left(\Phi(\sqrt{2 \log M}-1)\right)^{M - 1} \\
&  \leq \left( 1 - \frac{(\sqrt{2\log{M}} - 1)}{1 + \left(\sqrt{2\log{M}} - 1\right)^2}
\frac{\exp{\left(\sqrt{2\log{M}}\right)}}{M \sqrt{2\pi e}}\right)^{M - 1} \\
& = \Bigg[ 1 - \left(\frac{(\sqrt{2\log{M}} - 1)^2}{1 + \left(\sqrt{2\log{M}} - 1\right)^2}\right) 
\left(\frac{(M -1) \sqrt{2\log{M}}}{M (\sqrt{2\log{M}} - 1)}\right)  \\
& \hspace{0.5in}  \left(\frac{\exp{\left( \sqrt{2\log{M}}\right)}}{(M -1) \sqrt{2\log{M}} \sqrt{2\pi e}}\right)\Bigg]^{M - 1} \\
& \stackrel{(a)}{\leq} \left( 1 -  \frac{1}{(M-1)} \frac{\exp (\sqrt{2\log{M}})}{\sqrt{2\log{M}} \; \sqrt{4\pi e}} \right)^{M - 1} \\
& \stackrel{(b)}{\leq} \exp\left( - \; \frac{\exp (\sqrt{2\log{M}})}{\sqrt{2\log{M}} \; \sqrt{4\pi e}} \right).
\end{split}
\label{eq:PhiM_bound}
\ee
In the above, $(a)$ holds for large enough $M$, and $(b)$ is obtained using $1+x < e^x$.
Using this bound in \eqref{eq:I2_UB}  yields
\be
I_2 \leq \exp \left[ - \; \dfrac{\exp (\sqrt{2\log{M}})}{\sqrt{2\log{M}} \; \sqrt{4\pi e}}  + \log M + \sqrt{2\log{M}} \right].
\ee
Since the first term of the exponent dominates as $M$ grows large, the claim is proved.
\end{IEEEproof}

Finally we bound $I_3$ as follows.
\be
\begin{split}
I_3 & = \int_{\sqrt{2\log{M}} - 1}^{\sqrt{2\log{M}}}\exp\left(-t\left(\dfrac{z}{\sqrt{2\log{M}}} - 1\right)\right)\\
& \hspace{0.9in} \cdot M (\Phi(z))^{M - 1} \phi(z)dz \\
&\stackrel{(a)}{=} \int_0^1 \exp(\sqrt{2 \log M} \; u) \;  M (\Phi(\sqrt{2 \log M} - u))^{M - 1} \\
& \hspace{0.5in} \cdot \phi(\sqrt{2 \log M} - u) \, du \\
&{=} \int_0^1 \exp(\sqrt{2 \log M} \; u) \;  (\Phi(\sqrt{2 \log M} - u))^{M - 1} \\
& \hspace{0.35in} \cdot  \dfrac{\exp(\sqrt{2 \log M} \; u) \exp(-u^2/2)}{\sqrt{2\pi}} du \\
& \stackrel{(b)}{\leq} \max_{u \in [0,1]} g(u)
\end{split}
\label{eq:I3_ub0}
\ee
where
\[
g(u) =  \frac{1}{\sqrt{2\pi}} \exp(2 \sqrt{2 \log M} \; u) (\Phi(\sqrt{2 \log M} - u))^{M - 1}.
\]
In \eqref{eq:I3_ub0}, $(a)$ is obtained using the change of variable \[ z=\sqrt{2 \log M} - u \] and $(b)$ by bounding $\exp(-u^2/2)$ by $1$.

Using the upper bound \eqref{eq:Phi_bound} for $\Phi$ and steps similar to \eqref{eq:PhiM_bound}, we obtain
\be
\begin{split}
g(u) &  \leq \dfrac{1}{\sqrt{2 \pi}} \exp ( 2u \sqrt{2 \log M}) \\
&\quad  \cdot \exp\left[ - \;\dfrac{(1-\delta_M)}{\sqrt{4 \pi e \log M}} \exp (\sqrt{2 \log M} u) \right]
\end{split}
\label{eq:gu_UB}
\ee
where $\delta_M >0$ is a generic constant that  goes to $0$ as $M \to \infty$. It can be verified that the maximum of the right side of \eqref{eq:gu_UB} for $u \in [0,1]$ is attained at
\[ u = \frac{\log (16 \pi e \log M)}{2 \sqrt{2 \log M}} (1 + \delta_M) \] and from \eqref{eq:I3_ub0}, the maximum value is a upper bound for $I_3$. We thus obtain
\be
I_3 \leq \frac{8 \sqrt{2 \pi} (1+ \delta_M)}{e} \log M.
\label{eq:I3_ub1}
\ee

Using \eqref{eq:final_boundI0}, \eqref{eq:I1_bound}, Claim $1$ and \eqref{eq:I3_ub1} in \eqref{eq:int_split}, we conclude that
\be
\expec[\exp(2 \log M\; \abs{\epsilon_1})] \leq  1 + \frac{8 \sqrt{2 \pi} (1+ \delta_M)}{e} \log M \leq 8 \log M
\ee
for sufficiently large $M$. ($\delta_M$ is a generic constant that goes to $0$ as $M \to \infty$.) Using this in \eqref{eq:chernoff0}, we obtain
\be P\left(\sum_{i=1}^{L}\frac{\abs{\epsilon_i}}{L} > \delta\right) \leq \left( \frac{M^{2\delta} }{ 8 \log M} \right)^{-L}.\ee

\section*{Acknowledgement}
The authors would like to thank A. Barron and A. Joseph for several insightful discussions, and A. Greig for his help with  simulations. They would also like thank the anonymous reviewers for their detailed comments which led to an improved paper.

\IEEEtriggeratref{27}

\end{document}